\documentclass[11pt]{article}
\RequirePackage{latexsym}
\RequirePackage{graphicx}
\RequirePackage{amssymb}
\RequirePackage{setspace}

\usepackage{txfonts}
\usepackage[font=normalsize]{caption}
\usepackage{hyperref}
\usepackage{float}
\usepackage{wrapfig}
\usepackage[round,semicolon,authoryear]{natbib}

\topmargin -0.8in\graphicspath{ {figures/} }
\usepackage{array}
\newcommand{\apj}{    {\it Astrophys. J.}}
\newcommand{\apjl}{   {\it Astrophys. J. Lett.}}

\newcommand{\solphys}{{\it Solar Phys.}}

\chardef\us=`\_

\usepackage{color}
\usepackage{array}
\usepackage{url}
\usepackage{hyperref}
\usepackage{appendix}
\usepackage[dvipsnames]{xcolor}

\begin{document}

\definecolor{grey}{rgb}{0.5,0.5,0.5}
\definecolor{gold}{rgb}{0,0,0}
\definecolor{blue}{rgb}{0,0,0}
\definecolor{red}{rgb}{0,0,0}

     \begin{figure}
     \vskip 0.5in
     \begin{center}
     \includegraphics[scale=0.8]{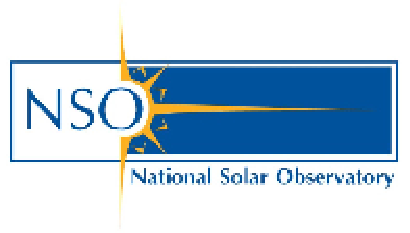}
     \end{center}
     \end{figure}

 \begin{center}
  \vskip 1in
{\bf\LARGE Anomaly Detection for GONG Doppler Imagery Using a Binary Classification Neural Network}

    \vskip 1in
{\Large Mitchell Creelman, Kiran Jain, Niles Oien and Thomas M. Wentzel}\\
\vskip 0.5in
 {\it\large NISP, National Solar Observatory, Boulder, CO 80303, USA}
\end{center}

\begin{center}
Technical Report No. {NSO/NISP-2024-001}
\end{center}

\vskip 0.5in

\noindent\rule{\linewidth}{0.5mm}
\rule{\linewidth}{0.5mm}

\begin{abstract}
        \addcontentsline{toc}{subsection}{\bf Abstract}
 
 One of the products of the National Solar Observatory's Integrated Synoptic Program (NISP) is the farside seismic map which shows the magnetic activity on the unobserved side of the Sun. The production of these rudimentary maps began in 2006, and they have since proven to be a valuable tool in tracking solar activity which cannot be directly observed from the earth's surface. The continuous tracking of solar active regions allows space weather forecasters to monitor critical solar events which may have larger economic and societal impacts here on Earth. In an effort to improve these maps, several steps are underway through the Windows on the Universe project (WoU) funded by the NSF. One of these steps is to improve the quality assurance measures for the images collected at individual sites throughout the GONG network and is used to develop the farside maps. To this end, we have designed a binary classification neural network to determine which of these site images should and should not be included in the farside pipeline that produces the end product maps. This convolutional neural network is a highly effective and computationally efficient method of significantly improving the quality of the farside maps currently produced by the NISP program.  
    
\end{abstract}

\pagebreak

    \tableofcontents

\pagebreak

\section[{Background}]{{Background}}
\label{Introduction}
\begin{figure} [h]
        \begin{center}
          \includegraphics[width=0.7\textwidth]{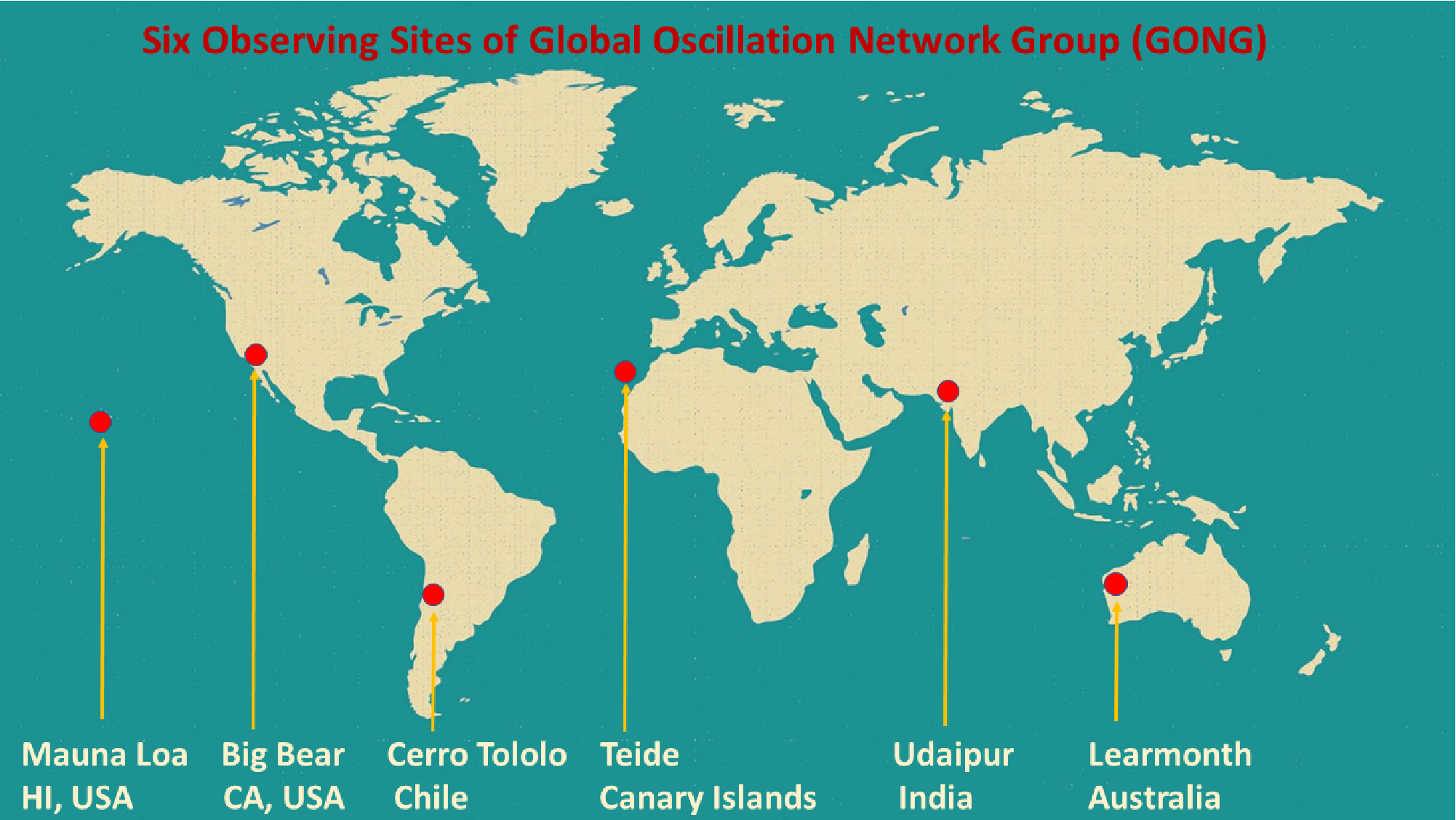}
            \caption{\it Global distribution of all six GONG network observatories.}
            \label{sites}
        \end{center}
    \end{figure}

    The Global Oscillation Network Group (GONG) is a network of ground-based solar observatories run by the National Solar Observatory's Integrated Synoptic Program (NISP) which is aimed at collecting uninterrupted solar observations \citep{Harveyetal1996,hill2018}. The network, established in 1995, is comprised of six observatories whose locations were chosen in order to create a redundant system capable of collecting solar observations with a mean duty cycle (DC) of over 90\%.  After a careful survey, observatories were established at Learmonth, Australia (\texttt{le}), Udaipur, India (\texttt{ud}), El Teide, in the Canary Islands (\texttt{td}), Cerro Tololo, Chile (\texttt{ct}), Big Bear, California (\texttt{bb}), and  Mauna Loa, Hawaii (\texttt{ml})  \citep{Hilletal1994}. The location of these observatories, shown in Figure~\ref{sites}, were carefully chosen to provide sufficient latitudinal and longitudinal coverage for consistent observational overlap between adjacent sites. A recent study by \cite{jain2021dutycycle} shows that the network is capable of collecting observations from as many as four sites simultaneously depending on the time of year. 

    While  multiple data products are derived from the observations of current GONG network\footnote{\url{https://gong2.nso.edu/archive/patch.pl}}, this report is primarily focused on the line-of-site low-resolution Dopplergrams which are produced and used to track the magnetic activity on the farside of the Sun. These Dopplergrams, or \texttt{fqi} images, are 215 $\times$ 215 pixel images which are produced from the high-resolution images collected at individual sites. The individual site minute-resolution Dopplergrams are then merged together to form a nearly continuous series of images. If there are no images for a given minute,  the produced Dopplergram is left empty. The fraction of 24\,hr day for which  the observations are available is known as the  
    duty cycle (DC), with a range between 0 and 1. The Dopplergrams are then used to determine phase shift values that are utilized to create helioseismic maps of the farside produced by the NISP program \citep{Lindsey_Braun2000,Braun2001,IGH_2007,IGH_2010}.   \\
    
    The phase shift maps are employed for a wide variety of space weather forecasting applications \citep[e.g.,][]{jain2022seismic}. In particular, they serve to help monitor the presence of known solar active regions, as well as the development of new regions on the farside of the Sun. Through developing improved farside helioseismic maps, we can improve the prediction and warning times for solar activity events, such as flares, coronal mass ejections (CMEs), and increase our understanding of the precursors for these events. One such example can be found in the sudden increase in the irradiance values during the early phase of the current solar cycle 25. While the frontside hemisphere saw a reduction in magnetic activity, a large active region which had emerged on the farside was found responsible for this sudden rise in the irradiance \citep{jain2021irradiance}. These farside seismic maps have also been used to create global magnetic index \citep[e.g.,][]{igh_2011,igh2014}, which is a useful proxy for the magnetic field models. 

\subsection[{How binned-down images are generated }]{{How binned-down images are generated}}
    \label{fqi-generation}
     To generate single-site binned-down 215 $\times$ 215 pixel \texttt{fqi} images, the near-real time 860 $\times$ 860 pixel full-resolution \texttt{vqi} images are  run through the Image Reduction and Analysis Facility (IRAF)/GONG data Reduction and Analysis Software Package (GRASP)'s \texttt{mkfsi} routine to fit a 2-d surface fit to the image and then subtract the obtained surface from the image to remove the solar rotation. A Gaussian filter is then applied before the images are down sampled to their final resolution. These \texttt{fqi} images from sites are transferred to \texttt{iQR} Keep at the NISP Data Center where they are processed by qr-filter to place them in the \texttt{oQR} Keep with altered header and added image statistics.  The next step is to merge all \texttt{oQR} single-site images for a given minute to create merged \texttt{mrfqi} image. During this process, the \texttt{oQR} single-site \texttt{fqi} images are rotated, flipped, and scaled to a standard radius prior to merging \citep[see][for general description]{Toner_2003, Hughes_2016}. \
        
        \begin{wrapfigure}{r}{0.55\textwidth}
                \vspace{-1.5cm}
                \includegraphics[width=0.55\textwidth]{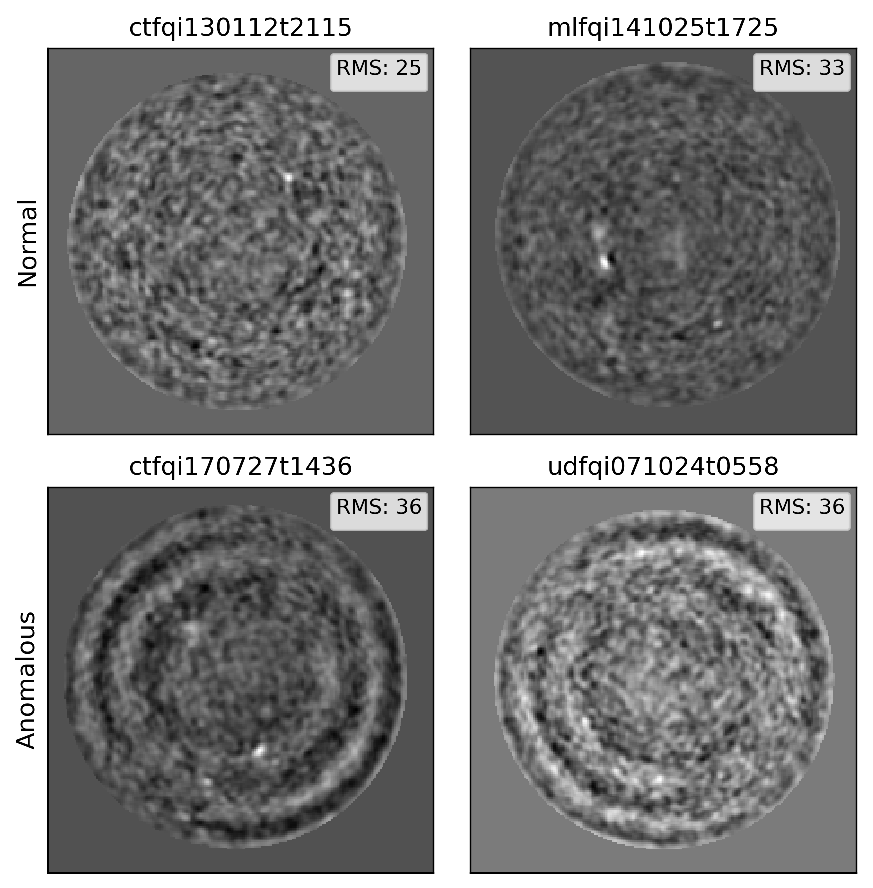}
                     \caption{\it Examples of normal (Top) and anomalous site images (Bottom) which would pass the current RMS site threshold.}
                    \label{FQI_samples}   
                \hspace{0.05cm}
                \vspace{-2cm}
        \end{wrapfigure}
        
\subsection[{What is the concern?}]{{What is the concern?}}  \label{Concern}
    The primary goal of this work is to develop a consistent and reliable methodology for excluding anomalous site images in the merged \texttt{fqi} images and thus the downstream farside data products.  Figure~\ref{FQI_samples} displays examples of the normal and erroneous images. The possible causes of anomalous images vary from malfunctioning calibration sequences, instrumental failures, or processing issues. The inclusion of anomalous Dopplergrams in the computation of farside seismic maps can have adverse consequences on the signal-to-noise ratio without affecting the duty cycle and rendering the final maps ineffective. \\
\vskip 0.5 cm

\section[{Image Quality Assurance and Control}]{Image Quality Assurance  and Quality Control}
    Implementing an effective quality assurance and quality control (QAQC) parameters for the site-specific \texttt{fqi} products is the overall focus of this report. Currently, the only  QAQC method in place is a thresholding of Root Mean Square (RMS) values. As will be covered in more detail later in this report, this methodology is largely ineffective, allowing many anomalous images to be included while excluding valuable data. The alternative proposed here is the implementation of a binary classification neural network for detection of anomalous \texttt{fqi} images. While not a perfect solution, such networks have been shown to be highly effective for anomaly identification tasks \citep{khan2021spectrogram}. 
    
\subsection[{Current QAQC Filter}]{Current QAQC Filter} \label{Current_QAQC}
    As mentioned, there is currently only one quality control parameter set on the \texttt{fqi} images generated at any individual site. This is a RMS filter with a threshold of 60. An inspection of the source code for the quick\_reduce script shows that this filter was implemented on July 23, 2014. Prior to this date, all images irrespective of their quality were included in the image-merge process and sometimes produced poor quality data products. This filter diverts all images with the RMS exceeding the threshold and stores them in the \texttt{Oubliete} Keep at sites. These images are then discarded after 41 days without being transferred to the NISP Data Center. This type of filter fails to include vital images during the high solar activity periods as well as to identify many of the anomalous images during the low-activity period. Figure~\ref{FQI_samples} displays examples of four site images with comparable RMS values where two of these images are erroneous.  Such a poor quality control measure has led us to explore the use of a binary classification neural network as an anomaly detection measure to replace the current filters.
        
\subsection[{Proposed QAQC Filter}]{Proposed QAQC Filter}
    For this application, we developed a filter using a simple binary classification network to label an \texttt{fqi} image as normal or anomalous. The developed filter seeks to prevent anomalous site-\texttt{fqi} images from being included in the \texttt{mrfqi} products by identifying anomalous site images  during the transition between the \texttt{iQR} and \texttt{oQR} keeps at the NISP Data Center. By excluding anomalous site-\texttt{fqi} images from the farside pipeline, we seek to improve the signal-to-noise ratio in the farside maps, thus providing a more reliable data product to the  users relying on the NISP data.\\
    
     The network was designed using the PyTorch Python package \citep{ketkar2021automatic}. PyTorch is a machine learning programming package that allows for a high degree of customization in the design and troubleshooting of the neural networks.
            
\section[{Training the Network}]{Training the Network}
        
\subsection[{Sampling}]{Sampling}  
    As mentioned above, the \texttt{fqi} products in the farside pipeline are velocity images of nearside of the Sun, and the phase shift values are calculated to determine the location of the active regions which are not directly observed from the Earth \citep{Lindsey_Braun2000}.\\  
 
    The efficacy of machine learning algorithms is highly dependent on the quality of the datasets that they are trained on. Though imbalanced datasets can be used to effectively train networks \citep{korkmaz2020deep}, balanced datasets can significantly improve the performance of machine learning algorithms \citep{divovic2021balancing}. However, this was complicated by the fact that we wanted to include a mix of normal and anomalous data within our training dataset. All of the data used in the training process were manually classified using an in-house tool developed by Kevin Shuman \citep{shuman2023good-bad-image-gui}. \\

         \begin{figure}[t]
            \centering
                \vspace{-1\intextsep}
                     \includegraphics[width=0.9\textwidth]{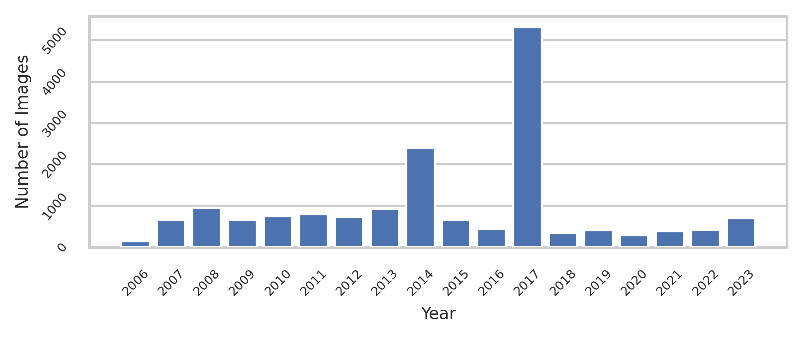}
                        \caption{\it Number of training dataset images per year.}
                         \label{training_yearly_hist}
                \vspace{-0.25cm}
            \end{figure}  
              
   To find anomalous data, we started with all images in the \texttt{fqi} dataset which had a variance of over 1600. Additionally, a random sample was taken of the oQR keep, with 5 random images taken from 20 days for a random 200 months within the lifespan of the GONG network. This left us with 110,593 high variance images, and 20,000 randomly sampled images. 21,268 of these images were then hand-sorted into normal or anomalous classes to be used to train the machine learning algorithm. \\
        \begin{wrapfigure}{l}{0.62\textwidth}
                \vspace{-0.5cm} 
                    \includegraphics[width=0.62\textwidth ]{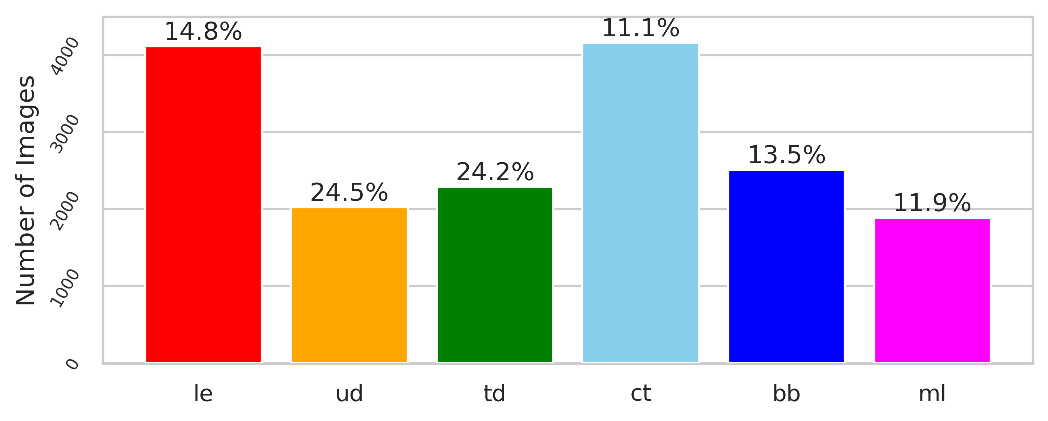}
                        \caption{\it Distribution of individual site contributions to the 17014 images in the training dataset. }
                    \label{training_hist}
            \vspace{-0.25cm}
        \end{wrapfigure}
    The year-wise distribution of the images used in training the network is shown in Figure~\ref{training_yearly_hist}. Using a training/testing split of 80\,\% to 20\,\%, 17,014  images were used to train the network while 4,254 images to test the performance of the classification network. The distribution of individual site images in the dataset is shown in Figure~\ref{training_hist}. \\ 
        
\begin{wrapfigure}{r}{0.62\textwidth}
       \vspace{-1\intextsep} 
       \vspace{-0.05cm}    
            \includegraphics[width=0.62\textwidth ]
            {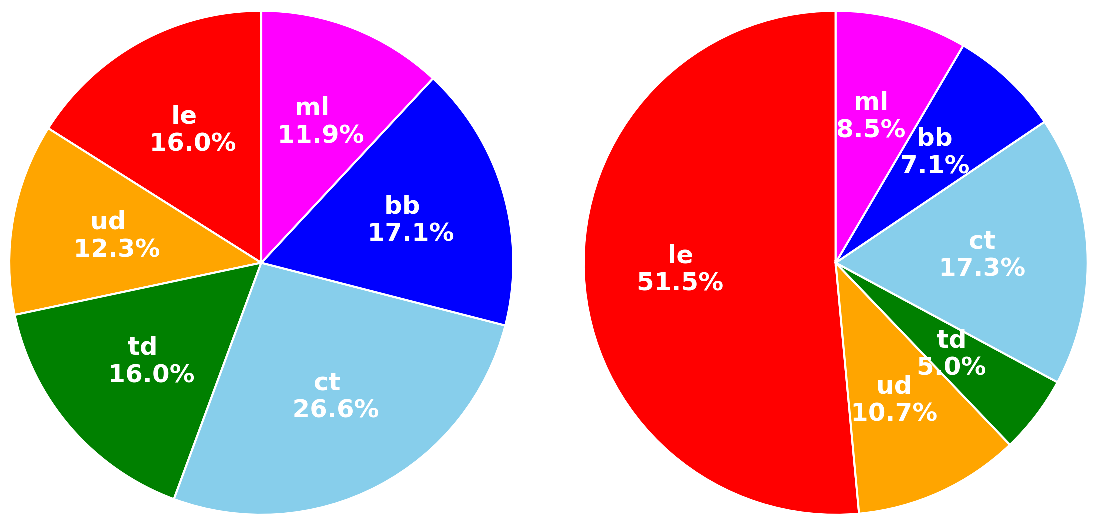}
        \caption{\it Distribution of 13098 normal (left) and 3916 anomalous (right) images by individual site for training dataset.}                              
        \label{training_pie}
       \vspace{-.8cm}
   \end{wrapfigure}

    A source of imbalance in the training dataset was the rate of site occurrences between normal and anomalous images. This is illustrated in Figure~\ref{training_pie} where \texttt{ct} is disproportionately represented in the training dataset at a rate nearly twice the other sites. While \texttt{ct} is the site with the largest sampling, it also appears as though \texttt{ct} and \texttt{le} produce a disproportionate number of anomalous images overall.   \\  
     It appears that \texttt{le}
   made up nearly half of the anomalous images present in the training dataset. While improved sampling strategies may result in a more even distribution of sites among normal images, such discrepancies in the anomalous images would most likely remain. Such differences in occurrence rate by site are certainly something to be monitored in the flagged images caught by the fqi filter during deployment. \\
          
\begin{figure}
        \centering
            \includegraphics[width=0.95\textwidth ]{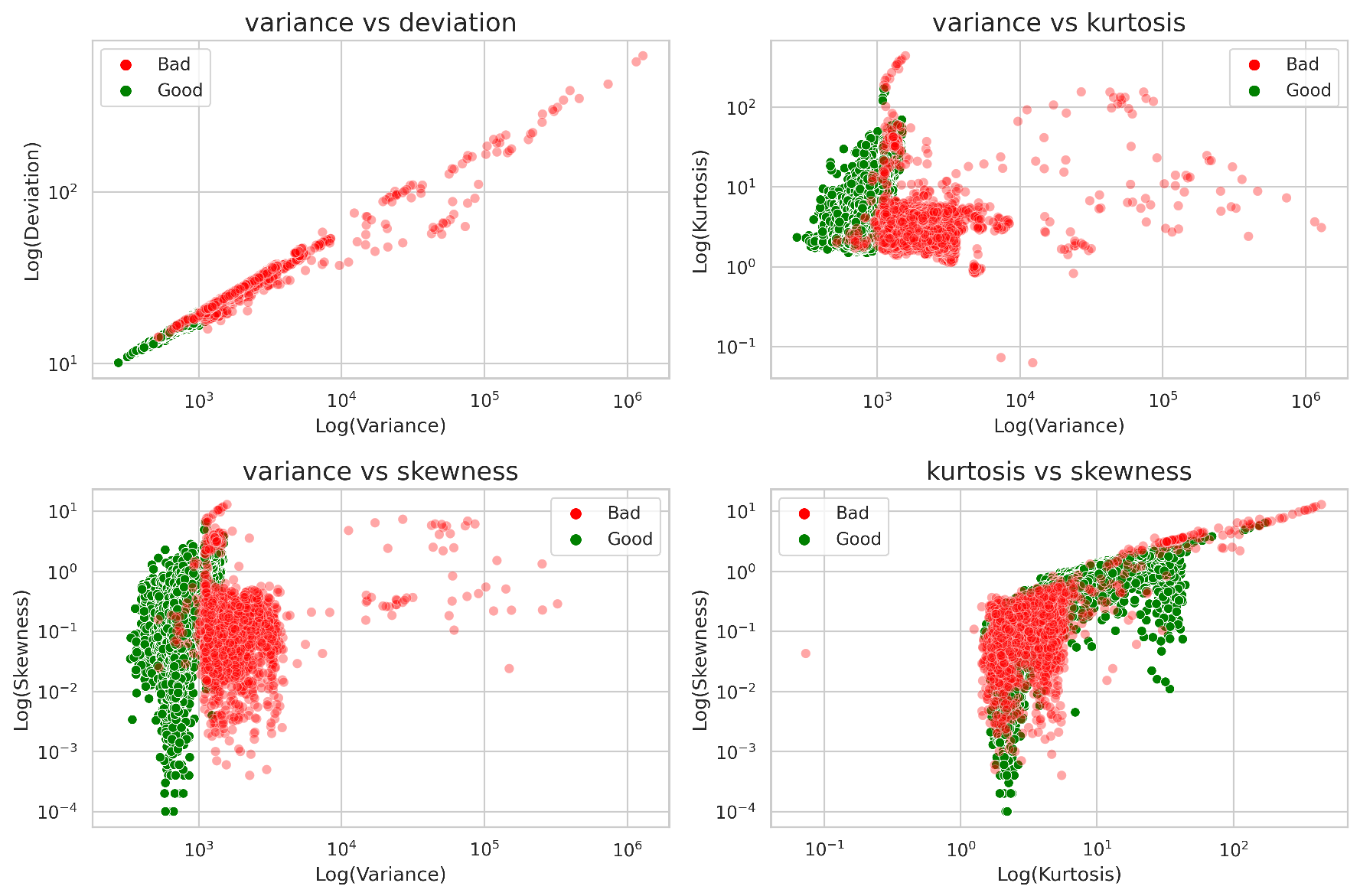}
                \caption{\it Statistics for training dataset highlighting the differences between normal (green) and anomalous (red) images.}
            \label{training_stats}
          \vskip -0.4cm
        \end{figure}

     Various statistical parameters for these images are plotted in Figure~\ref{training_stats}. Although the variance seems to be a good measure for identifying anomalous images, it becomes less effective during the high activity periods. In general, the erroneous images have higher variance as well as the RMS values relative to the normal images. The statistical overlap between the images during high-activity periods and the anomalous images often was seen in the database and it became one of the main motivations for the development of this filter. \\
   
    \begin{figure}[p]
        \centering
            \includegraphics[width=0.98\textwidth , height=20cm ]{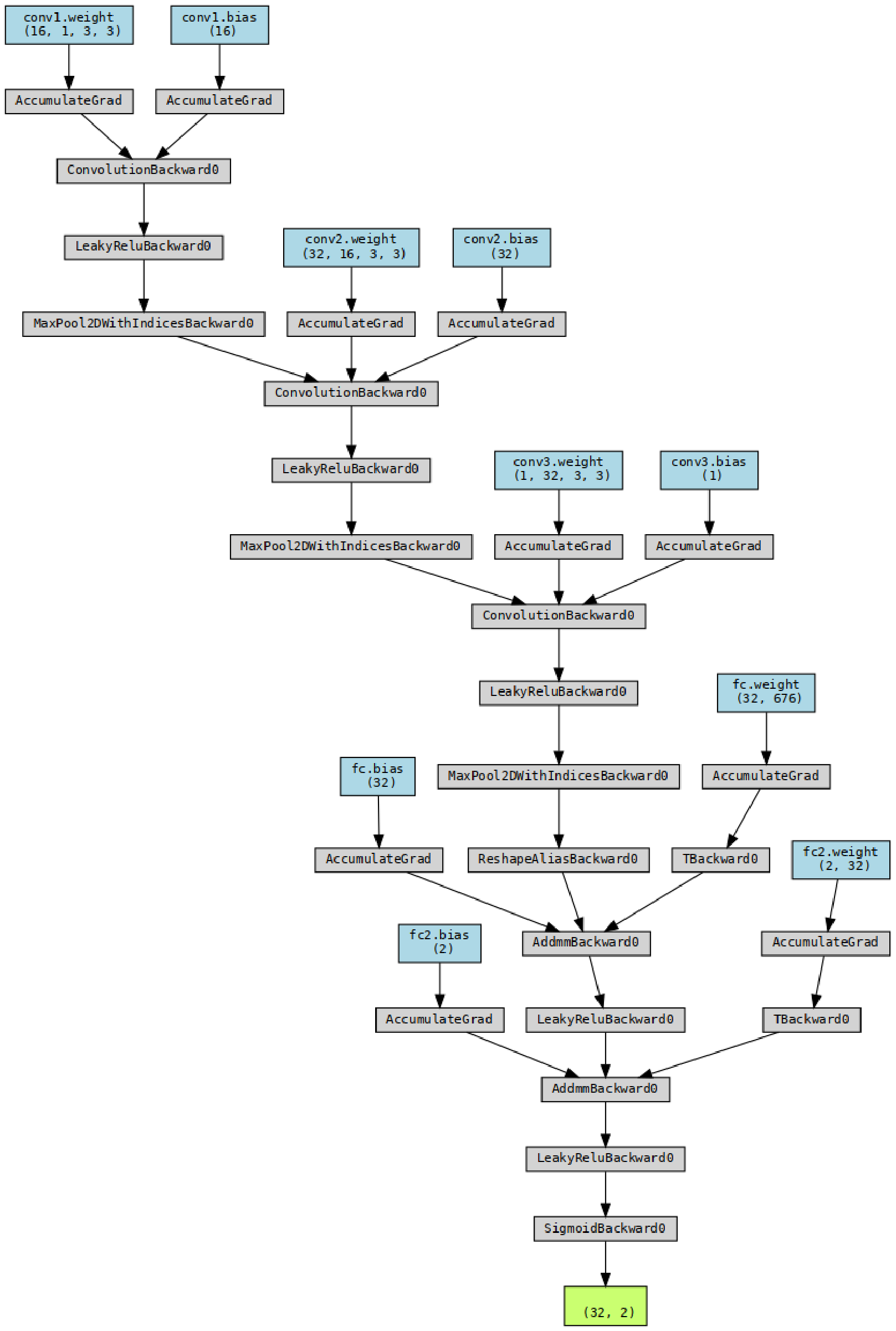}
                \caption{\it Outline of fqi Network design. This is based on a basic convolutional neural network which reduces the input image to a binary classification to determine whether the image is normal or anomalous.}
            \label{spottynet}
        \end{figure}
  
\subsection[{Network Design}]{Network Design}

    The design of the filter, outlined in Figure~\ref{spottynet}, is based on the neural network technique of image classification \citep{khalid2022deep,yal2023image}. However, while most classification algorithms look at a wide variety of classes \citep{yal2023image}, ours reduces the images into two bins: normal images and the anomalous images \citep{vailaya1998image}. The network design is simple, consisting of 3 convolutional layers, each followed by a leaky ReLU activation function and a max pooling. The leaky ReLU activation function was chosen for its better accuracy when compared to the commonly used ReLU activation function \citep{dubey2019comparative}. The network is ended with a sigmoid activation function due to its compatibility with more shallow networks \citep{khalid2022deep}. \\

     One of the most important choices in designing the network was the ability to weight the output classes in order to minimize the number of images falsely classified as anomalous. The reason for this concern is the statistical similarity between anomalous images and normal images during periods of high activity. 
     This means that if we allow the network to be more liberal in its tendency to classify images as anomalies, we run the risk of the network excluding large time-series of high-activity periods. To help prevent this, the network was designed to output a 2 $\times$ 1 vector containing scores for each possible class as discussed by \citet{khan2019weighted}. These scores range between a value of 0 and 1 for each class. 

\subsection[{Hyperparameters}]{Hyperparameters}
     While a variety of hyperparameters were explored for the training regiment, the final training regiment was set for 150 epochs with a batch size of 32, and a learning rate of 0.0008. The anomalous images were weighted at 1.6 while the normal images were weighted at 1.0 in order to minimize the occurrence of false anomaly classifications. \\

\section[{Performance Validation}]{Performance Validation}
    There were two experiments performed in order to validate the efficacy of the neural network filter. The first of these was a random sampling of 3250 site \texttt{fqi} images from the keep which were then classified by the filter and also manually checked. The second experiment was an analysis of 617 images were rejected at sites due to the RMS values above the threshold as discussed in Section~\ref{Current_QAQC} and had been moved to \texttt{Oubliete}.   

\subsection[{Test 1: Random Sample}]{Test 1: Random Sample}
    The first experiment was the utilization of the fqi network to classify 3250 randomly sampled images from the storage keep. Of the 3250 randomly sampled images, 1 was found to be corrupt and unusable, leaving 3249 images to be used for the experiment. These images were then passed through the fqi network, which classified the individual site images as either normal or anomalous. We also manually inspected the 3249 images to ground-truth the dataset. Of the 3249 images, 98.8\,\%  were correctly classified as normal images, and 0.8\,\%  were correctly classified as anomalous. In addition,  the network incorrectly classified remaining 0.4\,\% images. Of these 12 images, 10 were anomalous images which the network labeled as normal and 2 were normal images which the network misclassified as anomalous (Table~\ref{tab:Random_Sample_Matrix}). Overall, the network correctly identified normal images with an accuracy of 99.9\,\% and correctly identified anomalous images with an accuracy of 71.4\,\%  This is in line with the intended results, as the network's training regime was biased towards classifying images as normal in order to prevent discarding large amounts of good data due to the statistical similarities between anomalous and high-activity images as seen in the training dataset statistics (Figure~\ref{training_stats}).  
    
             \begin{table}[h]
                \centering
                    \caption{\it Confusion matrix for random sampling of fqi images from keep to validate the fqi Network. Of the 3250 images selected, one image was found to be corrupt and thus dropped from the overall sample. }
                \begin{tabular}{|c|c|c|}\hline
                     &  Actual Normal& Actual Anomaly\\\hline
                     Predicted Normal&  3212& 10\\\hline
                     Predicted Anomaly&  2& 25\\\hline
                \end{tabular}    
                \label{tab:Random_Sample_Matrix}
            \end{table}

 \subsection[{Test 2: RMS Rejections}]{Test 2: RMS Rejections}
    The second experiment was comparing the rejection performance of the fqi network to the current RMS threshold applied at sites before they are transferred to \texttt{iQR} and merged into \texttt{mrfqi} products. In this threshold, images are rejected if they have a RMS over 60. As mentioned above, the main drawback of this technique is that it can result in the rejection of images that are normal during high activity periods or allowing erroneous images to pass through during the low-activity periods. \\  

    For this experiment, we used 617 discarded site images that were stored in \texttt{Oubliete}. These images were manually checked to determine whether or not they had been correctly classified as anomalous by the thresholding methodology.  Of these 617 images, 91.6\,\% of these images were incorrectly discarded and contained valuable data, while 8.4\,\% of the images were correctly discarded for containing artifacts (Table~\ref{tab:RMS_Matrices}). It should be noted that it did not  determine the performance of the RMS threshold on images which were labeled as normal and this would have required the manually scanning of all data produced by the GONG network, which is outside of the current scope of this project.  \\
        
    The next step was to pass all discarded images through the newly developed fqi network. The network labeled 92.1\,\% of the images as normal and labeled 7.9\,\% of the images as anomalous. The normal image predictions had an accuracy of 99.6\,\%, while the anomalous predictions had an accuracy of 90.4\,\%. This gives the network an overall accuracy of 98.7\,\% with a false positive rate of 0.4\,\%, and a false negative rate of 9.6\,\% (Table~\ref{tab:RMS_Matrices}). \\
        
    The comparison is stark. Next to the 8.4\,\% accuracy of the RMS threshold, the neural network showed an overall accuracy of 98.9\,\%. Thus, we can say that implementing it as a site-based fqi filter would provide a sizable improvement in the quality of \texttt{mrfqi} images and by extension the farside seismic maps produced by the GONG network.   \\

            \begin{table}[h]            
                \vskip -0.25 cm
                \centering
                \caption{\it Test result comparison for RMS filtering (left) and the fqi network (right). The normal predictions for the thresholding method are excluded due to the volume of images, placing the category outside the scope of this study. }           
                \begin{tabular}{|c|c|c|c|c|c|c|}\cline{0-2}\cline{5-7}
                
                     Threshold&  Actual Normal&  Actual Anom.&  &  fqi Net&  Actual Norm.& Actual Anom. \\\cline{0-2}\cline{5-7}
                     
                     Pred. Norm&  N/A&  N/A&  &  Pred. Norm&  563& 5\\\cline{0-2}\cline{5-7}
    
                     Pred. Anom&  565&  52&  &  Pred. Anom&  2& 47\\\cline{0-2}\cline{5-7}
                \end{tabular}
                \label{tab:RMS_Matrices}
                \vskip -0.2 cm
            \end{table}
\section[{Effects of Implementation}]{Effects of Implementation}  
        
    There are two main areas of concern for implementing the fqi filter. The first is what is the effect it has on duty cycle. If the filter proves to take large sections of data, and reduce the duty cycle of the GONG network to below the generally accepted usability threshold of 0.8 or 80\,\%, then we would not consider it adequate as a QAQC measure.  The second area of concern is the effect it has on the farside maps. If the application of the filter results in significantly lower quality of the farside maps, it would not be considered adequate as a QAQC measure.\\
    
\begin{figure}[H]
            \centering
                \vspace{-0.5cm}
                \includegraphics[width = 0.82\textwidth]{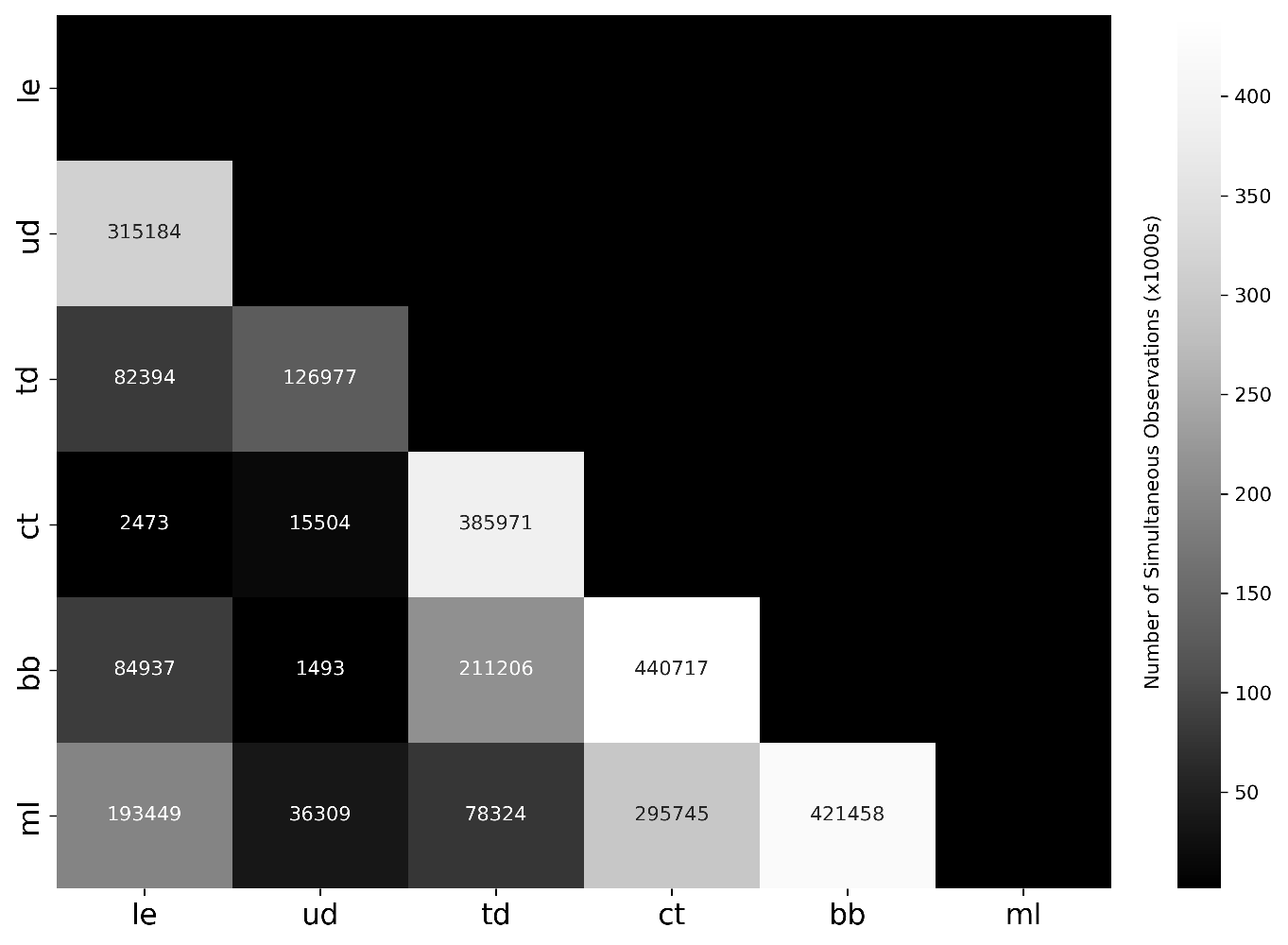}
                    \caption{\it Heatmap showing the frequency of simultaneous observations for all site combinations for the testing period.}
                \label{site_combos}    
            \vspace{-1.25cm}
        \end{figure}

    In order to determine the effects of the developed fqi filter on duty cycle, we examined all images collected between 2010-01-01 and 2014-12-31 for, a total of 1826 days, or 2,629,440 minutes. It is important to mention that our data set include only those observations that were transferred to the NISP Data Center. All rejected site images were lost due to an existing criterion discussed in Section~\ref{Current_QAQC}  and were unrecoverable.  Details of the contributions from various sites are given in Table~\ref{tab:simultaneous_obs}.  It is evident from the table that  all sites provide independent as well as overlapping observations.  
    Though there are significant overlaps between different sites,  the sites observed independently for about one-third of the total time. The maximum independent observations came from the \texttt{le} site. It is worth mentioning that erroneous images during the independent observations can have unfavorable effect on the duty cycle.  Figure~\ref{simul_obs} illustrates the distribution of multiple-site observations during the day. One can easily notice that the overlaps between multiple sites are not random but these follow a specific pattern depending upon the season and the time during the day.  The heatmap, shown in Figure~\ref{site_combos}, summarizes the number of overlapping observations between different pairs of sites.  Since the observations are often overlapped, it is important to quantify the effect, if any, of fqi filter on the overall network duty cycle. We also discuss the performance of each site during the test period. These effects are the concentration of the next two subsections of this report. \\ 
    
            \begin{table} 
            \centering
                   \caption{\it Rate of simultaneous observations from 2010-01-01 to 2014-12-31. This covered a period of 1826 days, or 2,629,400 minutes. The image counts for the individual sites show how often each site observed as a group of 0, 1, 2, 3, or 4 sites with the \% of time as part of a group in parentheses.} 
                   \vskip 0.1in
                \begin{tabular}{ccccccccc} \hline
                     \# of Sites   &  minutes    &  \texttt{le} &  \texttt{ud} &  \texttt{td} &  \texttt{ct} &  \texttt{bb} & \texttt{ml} \\ \hline 
                             0     &    153309   &       0      &         0    &          0   &       0      &         0    &    0        \\ 
                                   & (5.83\,\%) &              &              &              &               &             &             \\
                      \\
                              1    &   880548   &   288664     &  99175&  192205&  137100&  81304& 82100
            \\  
                                   &(33.49\,\%) &(33.81\,\%) &(19.05\,\%) &(24.05\,\%) & (15.57\,\%)& (10.14\,\%) &(11.07\,\%)\\
                      \\
                            2      &  1116679   & 452873&  347749&  374118&  391754&  326827& 340037\\
                                   &(42.47\,\%) &(53.05\,\%) &(66.80\,\%) &(46.81\,\%) & (44.50\,\%)& (40.78\,\%) &(45.85\,\%)\\
                      \\
                            3      &   432654   & 111027&  73256&  187757&  305556&  347117& 273249\\ 
                                   &(16.45\,\%) &(13.00\,\%) &(14.07\,\%) &(23.49\,\%) & (34.71\,\%)& (43.31\,\%) &(36.84\,\%)\\
            \\
                             4     &    46250   &  1170&  402&  45080&   45848&   46250& 46250\\
                                   &(1.76\,\%)  &(0.14\,\%) &(0.08\,\%) &(5.64\,\%) & (5.21\,\%)& (5.77\,\%) &(6.24\,\%)\\
                     \hline
                \end{tabular}
                    \label{tab:simultaneous_obs}                
            \end{table}      

   In the next section, we look at the effects of filtering out anomalous \texttt{fqi} images on the  farside maps. There, we show multiple examples of farside maps where the duty cycle either remained unchanged, or only slightly reduced with the application of fqi filter, and examine the effect excluding anomalous images has on the signal-to-noise ratio of these developed maps.  
   
\subsection[{Impact on Network Duty Cycle}]{Impact on Network Duty Cycle}
\begin{figure}[H]
         \vskip 0.5 cm
            \centering
                \includegraphics[width=0.99\textwidth,height=14 cm]{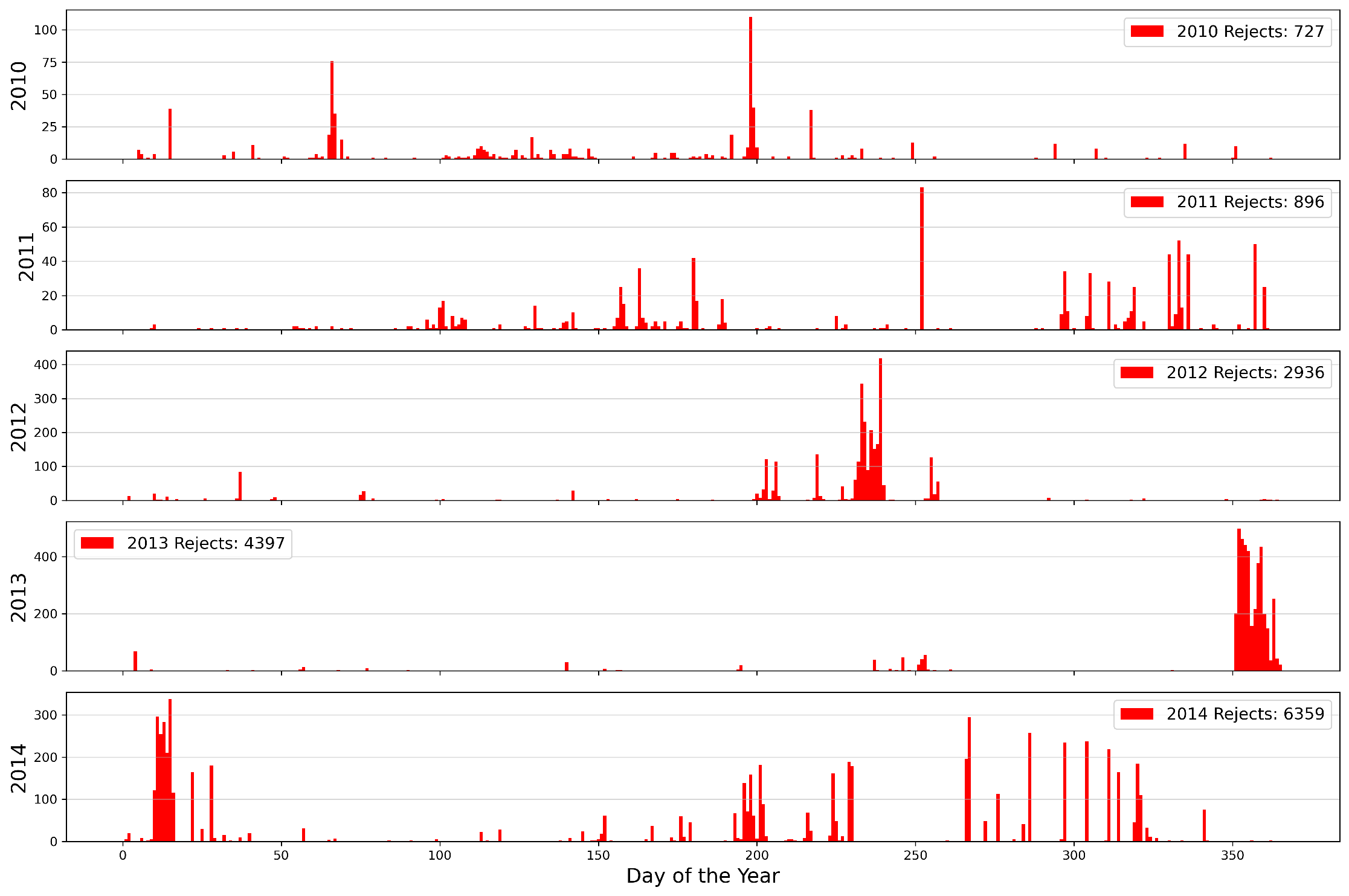}
                    \caption{\it Number of images rejected each day of the sample period. Note the high volume events for each year which make up periods of consistent anomaly production.}
                 \label{daily_rejects}
            \end{figure}

     The impact of the fqi filter on the network duty cycle is an important  concern for the implementation of this newly developed quality assurance tool as well as for the performance of the GONG network. If the network is too aggressive in its labeling of anomalous images, we risk losing large time series of good data. This is particularly difficult due to the overlap in statistical similarities between anomalous and high-activity images (Figure \ref{training_stats}). However, if the network lets too many images pass through, then it will be ineffective as a QAQC measure. An analysis of the impacts of the fqi filter during our study period shows that it has been able to remove a significant number of anomalous images from the test dataset without seriously impacting the overall duty cycle of the GONG network. We summarize total number of rejected images at each site in  Table~\ref{tab:site_contrib} and Figure~\ref{daily_rejects} displays the daily distribution of these rejected site images. We discuss two types of images in the following sections; the {\it unfiltered} and {\it filtered} images are the images prior and after passing through the proposed fqi filter, respectively. \\ 
     
            \begin{figure}[H]
            \centering
                \includegraphics[width = 0.95\textwidth]{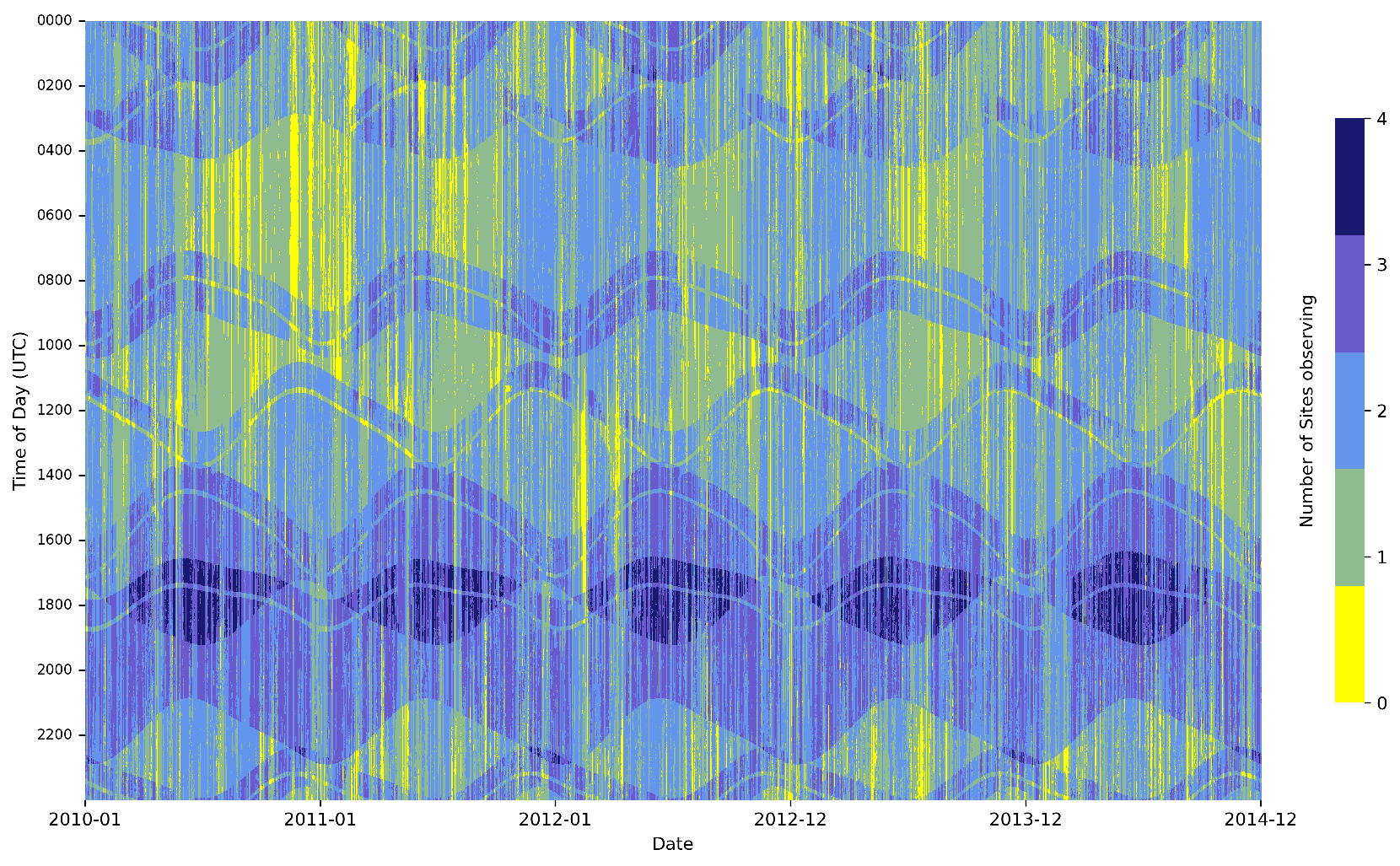}
                    \caption{\it Minute-resolution figure showing the number of sites observed simultaneously during the day for the entire study period. Note that the temporal and seasonal patterns are clearly visible through density fluctuations.}
                \label{simul_obs}
            \end{figure}
                        
          \begin{table}[H]
              \centering
                    \caption{\it Total number of images contributed by each site and the number of rejected images by the fqi filter. The \% values indicate the percentage of all rejected images contributed from each site.}
              \begin{tabular}{ccccccc}\hline      
                   Site&  \multicolumn{2}{c}{Total Images (Unfiltered)}&  &\multicolumn{3}{c}{Rejected Images (Filtered)}\\ 
                   \cline{2-3} \cline{5-7}
                 &  Total         \#         &  \% & & Total \#&  \% of site Images& \% of All Rejections\\ \hline
                   \texttt{le}&  853734&  18.50& & 624&  0.07 & 4.10\\
                   \texttt{ud}&  520582&  11.30& & 1454&  0.28 & 9.50\\
                   \texttt{td}&  799160&  17.40& & 6510&  0.81 & 42.50\\
                   \texttt{ct}&  880258&  19.10& & 328&  0.04 & 2.10\\
                   \texttt{bb}&  801498&  17.40& & 5974&  0.75 & 39.00\\
                   \texttt{ml}&  741636&  16.10& & 425&  0.06 & 2.80\\ 
                   \\
         Total:& 4596868& 100.00 & &  15315& 0.33 &100.00\\
         \hline
              \end{tabular}
              \label{tab:site_contrib}
          \end{table}      

      \begin{wrapfigure}{r}{0.72\textwidth}
           \vspace{-0.1cm}
            \centering
            \includegraphics[width=0.72\textwidth  ]{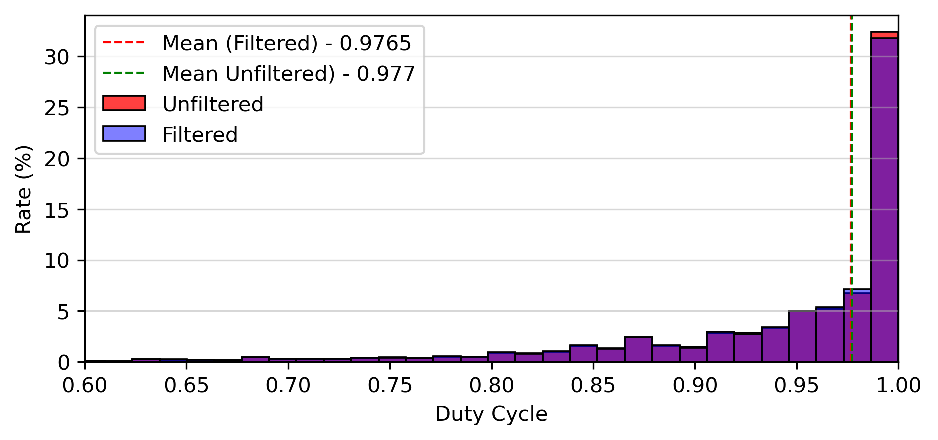}
                \caption{\it Changes in network duty cycle due to fqi network implementation on the test dataset. The dashed lines show mean duty cycles for filtered and unfiltered images.} 
                \label{dc_hist}
            \vspace{-0.5cm}
       \end{wrapfigure}
   
\begin{wrapfigure}{l}{0.72\textwidth}
        \centering
            \vspace{-1\intextsep}
            \includegraphics[width=0.72\textwidth  ]{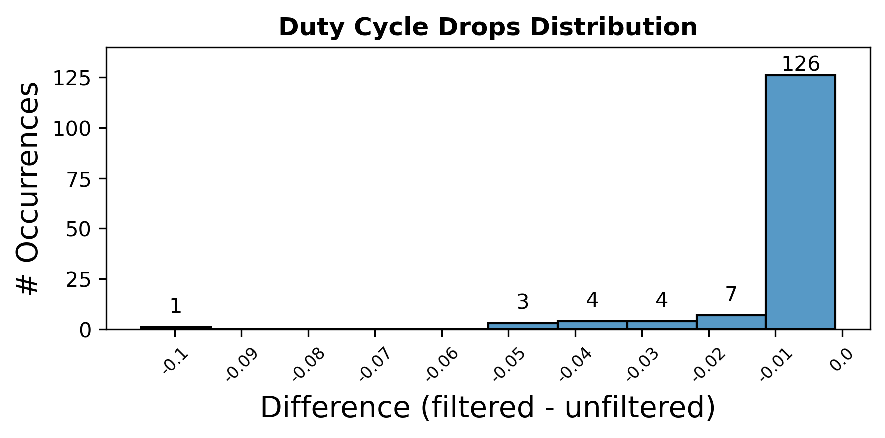}
                \caption{\it Drops in duty cycle due to application of the fqi filter excluded 15,315 of the 4,596,868 images, affecting 145 days. }
                \vspace{-0.3cm}
            \label{dc_drops}
       \end{wrapfigure}

\begin{figure}[t]
            \centering
            \includegraphics[width = \textwidth]{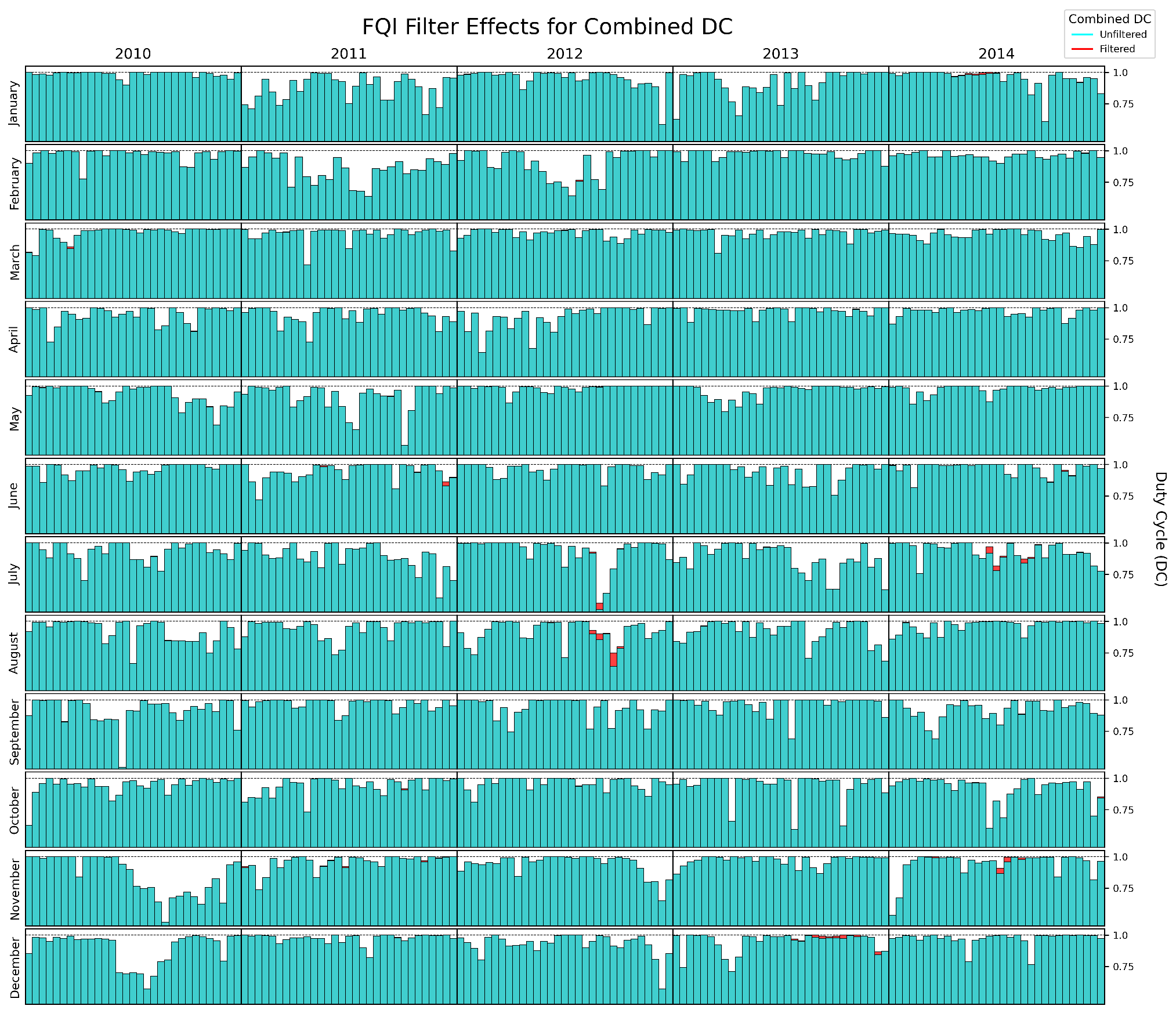}
                \caption{\it Daily duty cycle for the network-merged images during the testing period from 2010 through 2014. The effects of the fqi filter are shown in red and the lower y limit shown is 0.45. }
                \label{comb_dc}
            \vspace{0.3cm}
        \end{figure}

     In short, we find that the overall impacts on the GONG network's duty cycle were minimal after passing all site images through the fqi filter. From Table~\ref{tab:site_contrib}, it can be seen that of the 4,596,868 images in the test dataset, the network excluded only 0.33\,\% of the total dataset. The daily distribution of rejected site images and the yearly total rejections for the testing period, exhibited in  Figure~\ref{daily_rejects}, clearly suggests the sporadic nature of these rejections 
     and the occurrence of these anomalous events is unpredictable. We notice that the maximum rejections happened during 2014. This resulted in the slight reduction of mean duty cycle  from 94.17\,\% to 94.12\,\%, and the median duty cycle from 97.70\,\% to 97.65\,\% (Figure~\ref{dc_hist}). Due to significant overlaps between different sites, there were only 145 days when the network duty cycle was affected by the exclusion of erroneous images. Of these days, as shown in Figure~\ref{dc_drops}, the duty cycles were reduced by 0 -- 1\,\%,  1 -- 2\,\%, 2 -- 4\,\% and 4 -- 5\,\%,  for 126, 7, 8 and 3 days, respectively. It is visible from Figure~\ref{comb_dc} that there was only one day when the duty cycle for the GONG network reduced by 10\,\%.\\
            
    The relatively minimal impact of the exclusion of bad site images on the network duty cycle is attributed to the distributed and redundant nature of the GONG network \citep{Hilletal1994}. While there are temporal coverage patterns of the overall network as shown in Figure~\ref{simul_obs} and also discussed in detail by \citet{jain2021dutycycle}, the network did not collect any observations for 5.7\,\% of the total minutes during our study period. Overall, total independent observations from all individual sites in the network were 33.4\,\% of the total examined minutes, the simultaneous observations by two, three and four sites were 42.7\,\%, 16.5\,\% and  1.6\,\%, respectively. It is worth mentioning that the network's duty cycle is most vulnerable to impact from the exclusion of site-images during the independent observing periods. These opportunities are especially prevalent during the period between July and October when the \texttt{ud} site is prone to be down due to the Indian monsoon or any other site being down due to instrument failures or other reasons \citep{jain2021dutycycle}. The specifics of which sites often observe at the same time will be covered in the site-by-site analysis in the next section.
 
\subsection[{Impact on Site DC}]{Impact on Site DC}     \label{site_impact}
 
\begin{figure} [t]
        \centering
        \vspace{-\intextsep}
        \includegraphics[width = 0.85\textwidth]{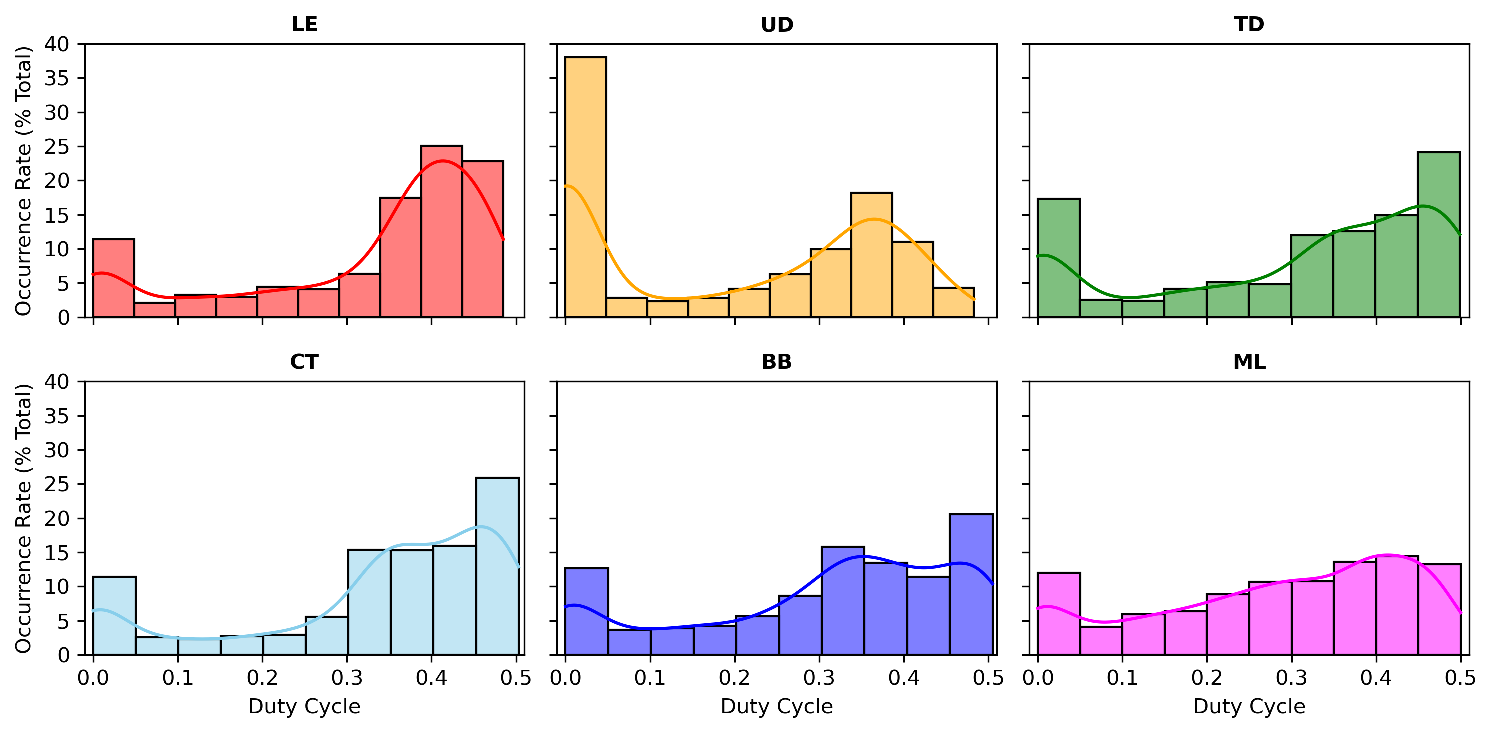}
        \vskip -0.15 in
            \caption{\it Histogram of the distributions of duty cycles for each site before the fqi  filter is applied. } 
        \label{site_dc_hist}
    \end{figure}

     While the potential impact of the fqi filter on the duty cycle of entire network is of obvious importance, we also investigate its effects on the duty cycles for individual sites. We display distribution of duty cycle at each site in Figure~\ref{site_dc_hist}. Aside from the obvious concern, that the overall resiliency of the network is dependent on its component sites, we also want to explore when and which sites produced anomalous images and what could be the probable reason. For reliably identifying anomalous image occurrences at each site in near-real time, this fqi filter may be proven useful for setting up a flag for an instrument malfunctioning if a large number of images are rejected. This information may be useful for the NISP operations team who may be able to identify and troubleshoot the regularly occurring problems in a timely manner. It should be noted that, due to the time discrepancy between data collection and when this test was conducted, any site images which have been excluded at the sites were not  included for the purpose of DC analysis. However, as the prior performance comparison showed, we can expect the results to be comparably favorable.   

        \begin{wrapfigure}{r}{0.53\textwidth}
             \vspace{-2.85\intextsep}
                \includegraphics[width = 0.53\textwidth , height = 14 cm] {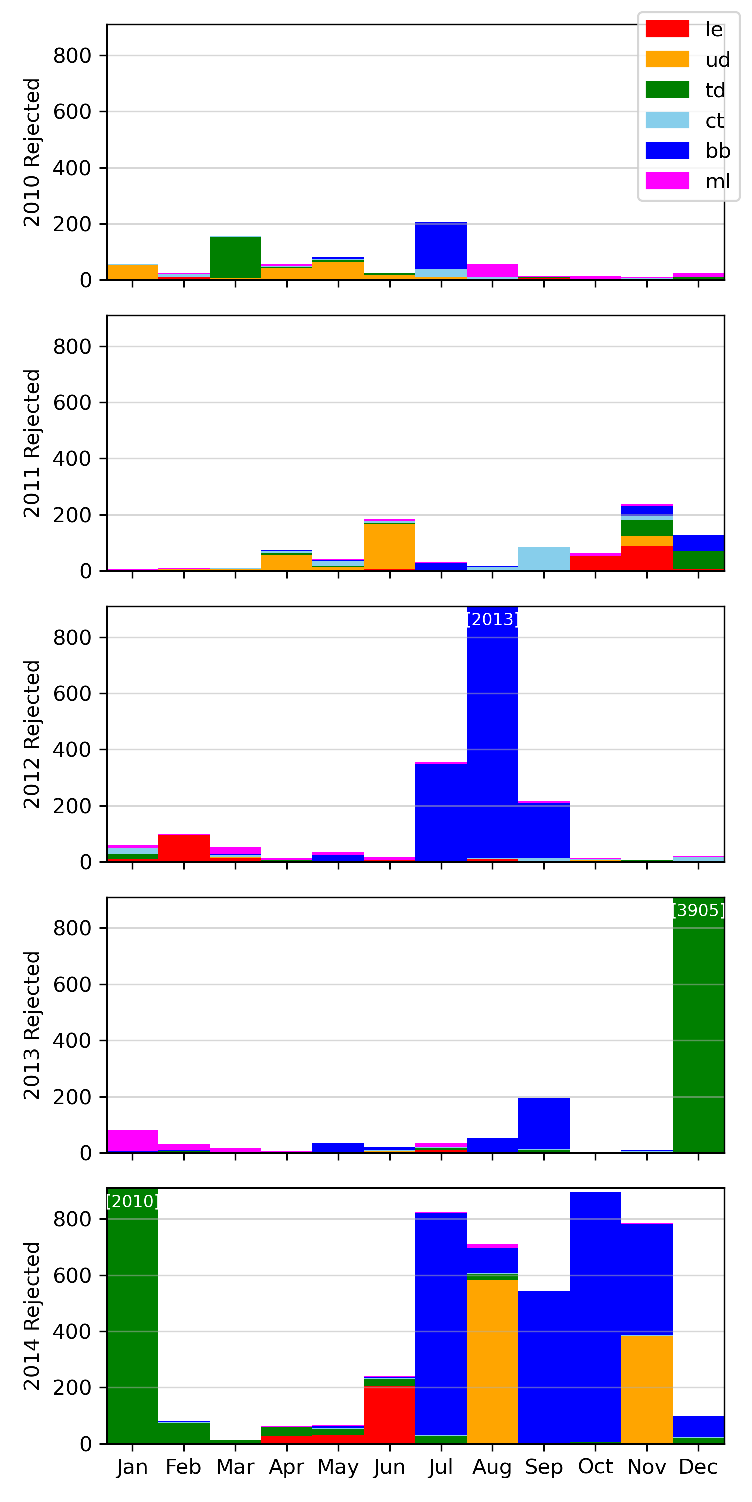}  
                 \caption{\it Images rejected by month for the sample period. Site contributions are shown in different colors. }
                \label{monthly_rejects}
            \vspace{-1.25cm}
        \end{wrapfigure}
           
 \subsubsection{Learmonth (\texttt{le})}
    Learmonth (\texttt{le}) had a mean daily duty cycle of 32.47\,\% for the testing period and as exhibited in Figure~\ref{site_dc_hist}, for the majority of its days, the duty cycle was between 0.30 and 0.50. The details of total observations and the rejected images are given in Table~\ref{tab:site_contrib}.  Overall it collected 853,734 minutes worth of observations, which makes  18.50\,\% of the total observations for the testing period. Of these total minutes observed by \texttt{le}, about one-third observation had no overlap with any other site and more than half were overlapped by one other site. Rest of the observations were gathered with two or three other sites (see Table \ref{tab:simultaneous_obs}). It is further evident from Figure \ref{site_combos} that the maximum overlap of \texttt{le} was with \texttt{ud}, followed by \texttt{ml}, \texttt{td}, \texttt{bb}, and \texttt{ct}, respectively.  
    
\begin{figure}[t]
         \centering
            \includegraphics[width = \textwidth,height=14.85 cm]{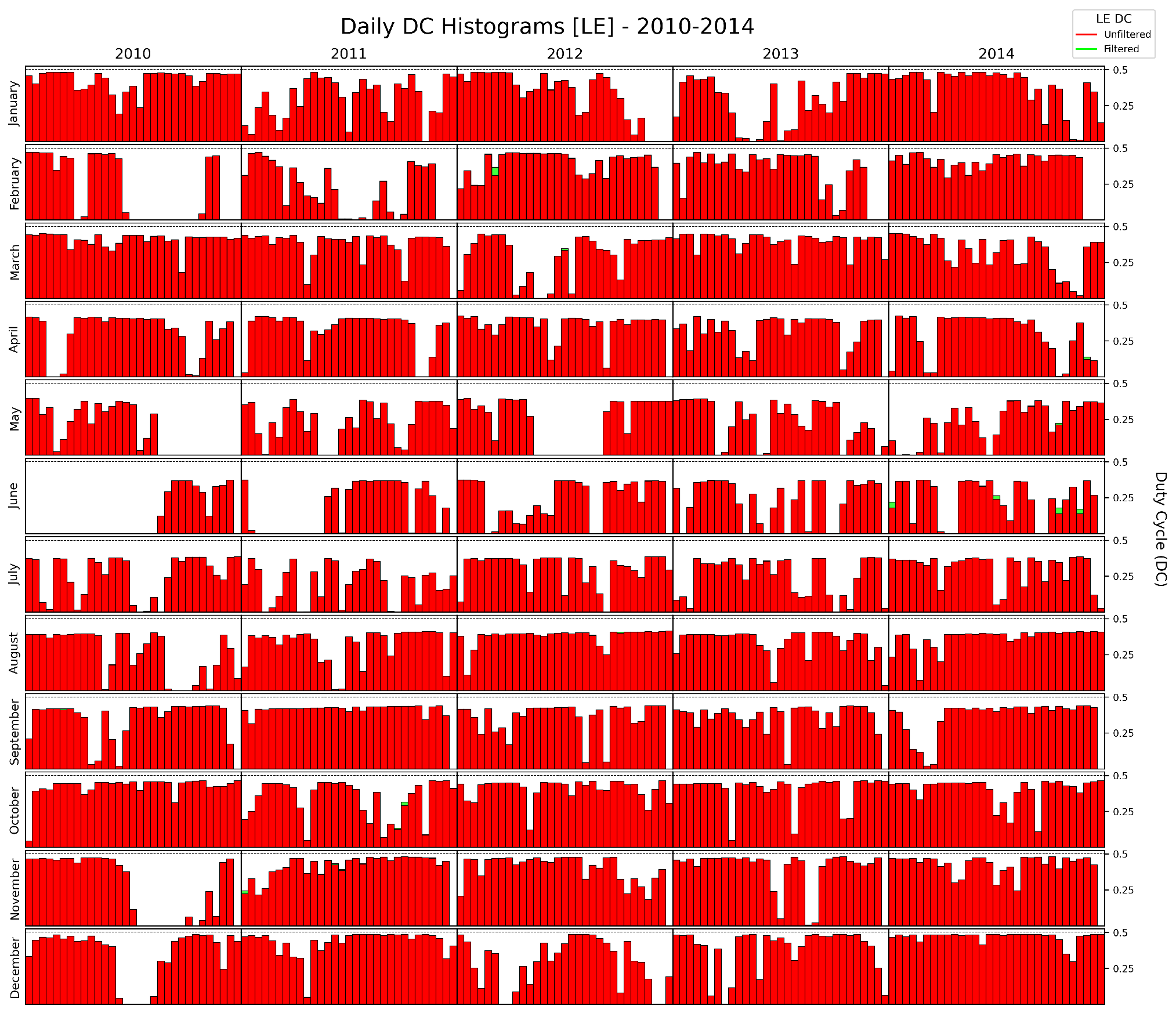}
            \caption{\it Daily duty cycle for the \texttt{le} site for the testing period from 2010 through 2014. The effects of the filter on duty cycle are shown in green. The major drops in duty cycles are seen on one day in February 2012 and several days in June 2014.}
          \label{dc_drop_le}
            \vskip -0.1 cm  
        \end{figure}

    From Table~\ref{tab:site_contrib}, it can be seen that of all the images collected by \texttt{le}, 624 (0.07\,\%) images were rejected by the fqi filter for being anomalous. This comprises 4.1\,\% of all the anomalous images found by the fqi network. Figure~\ref{monthly_rejects} provides number of the monthly rejections each year. We did not find any outstanding events contributing to anomalous images from the \texttt{le} site. This is also confirmed by a lack of visible drops in the daily duty cycle plotted in Figure \ref{dc_drop_le}.  Overall, the application of the fqi filter reduced the mean duty cycle of the \texttt{le} site from 32.47\,\% to 32.45\,\% (Table~\ref{tab:filtered_site_stats}) and did not discard all the observations for a given day.  
    

\subsubsection{Udaipur (\texttt{ud})}
\begin{figure}[t]
            \centering
                \includegraphics[width = \textwidth,height=14.85 cm]{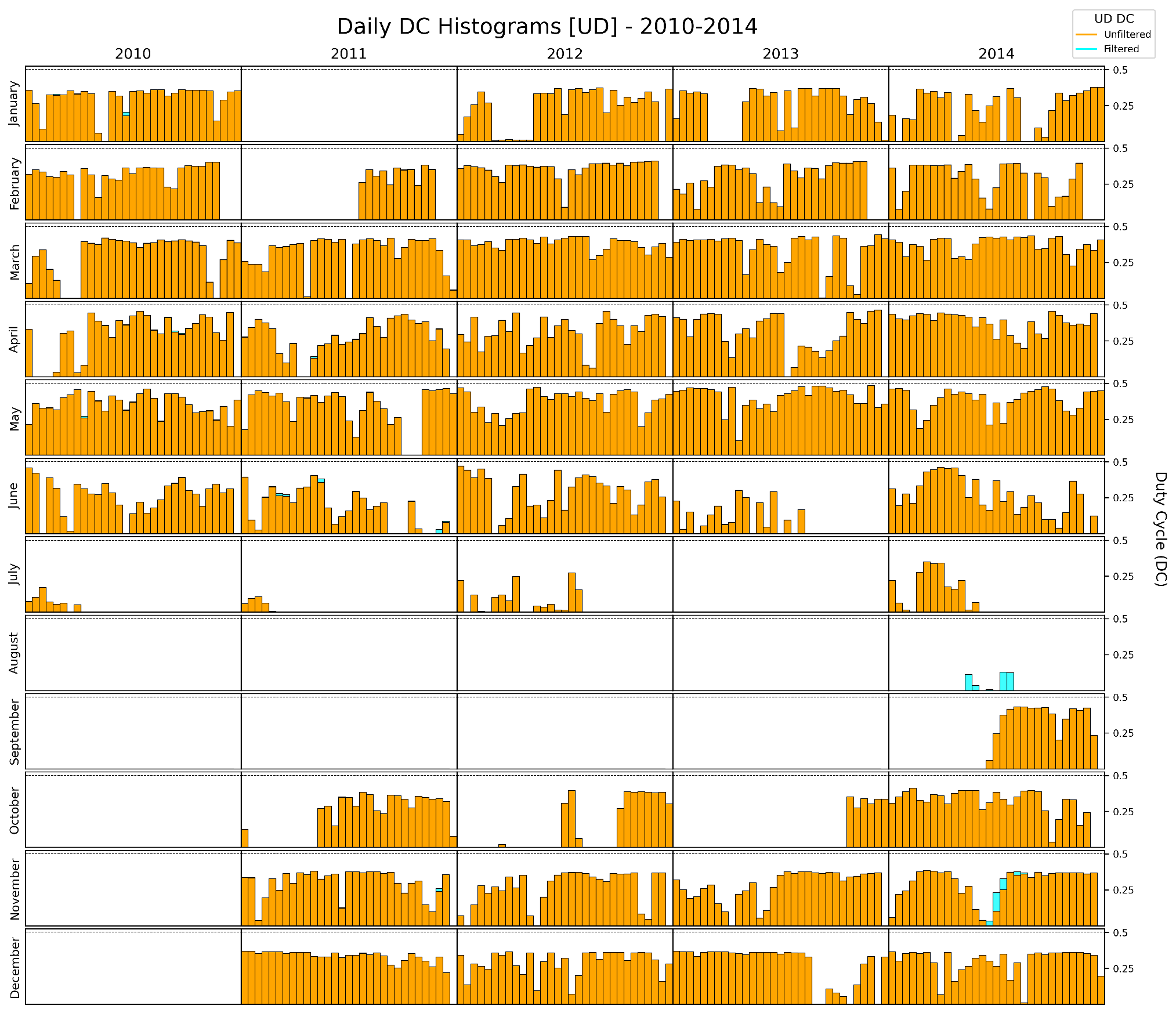}
                    \caption{\it Daily duty cycle for the \texttt{ud} site for the testing period from 2010 through 2014. The effects of the fqi filter on duty cycle are shown in cyan. The major drops in duty cycles are  apparent in August 2014 and November 2014. }
             \label{dc_drop_ud}
        \vskip -0.6 cm
        \end{figure}

    As illustrated in Figure~\ref{site_dc_hist}, while a majority of the non-zero days for the \texttt{ud} site had a duty cycle between 0.3 and 0.5, there are a disproportionate number of days when it did not contribute any observation to the GONG network. This is due to the annual monsoon season, which forces \texttt{ud} site to shutdown for a couple of months every year. These periods are clearly visible in Figure~\ref{dc_drop_ud}. In addition, there was a disruption which lasted for several months from mid 2010 to the beginning of 2011 due to a fire near the \texttt{ud} site. These long outages adversely affected the mean and median duty cycles of \texttt{ud} for the testing period of five years and the mean and median daily duty cycles were 19.80\% and 23.85\,\%, respectively, as summarized in Table~\ref{tab:filtered_site_stats}.  Of the 520,582 minutes that the \texttt{ud} site observed (Table~\ref{tab:site_contrib}),  19.05\,\% were independent without overlapping with any other site.  About two-third of \texttt{ud} observations were collected simultaneously with only one other site. There were relatively smaller number of observations that had overlaps with two or more other sites.  Details of these overlapping observations are given in Table~\ref{tab:simultaneous_obs}. As illustrated in Figure~\ref{site_combos}, the \texttt{ud} site most often observed in conjunction with two neighboring sites, i.e. \texttt{le} and \texttt{td} sites.  \\
    
\begin{table}[t]
    \centering
            \caption{\it Duty cycle statistics for filtered vs. unfiltered images for each of the GONG sites during the testing period of five years.   The percentage statistics are taken to 2 decimals places as many of the changes are too minute to be seen otherwise.}
         \vskip 0.1in         
    \begin{tabular}{ccccccccc}\hline
         Site&  Filter    &  Mean  & Median&  Standard   &  Maximum   & Minimum & \# of Days Dropped\\ 
             &            & DC [\%]&DC [\%] &  Deviation [\%] & DC [\%]  &  DC [\%]& to 0\,\% DC \\ \hline
         \texttt{le} &  Unfiltered &  32.47 & 38.20 &  14.86 &  48.50 & 0 &  \\
            &  Filtered   &  32.45 & 38.20 &  14.87 &  48.50 & 0 & 0 \\
              &    &   &  &   &   & \\          
         \texttt{ud} &  Unfiltered &  19.80 & 23.85 &  17.27 &  48.30 & 0 &  \\
            &  Filtered   &  19.74 & 23.85 &  17.29 &  48.30 & 0 & 6 \\ 
              &    &   &  &   &   & \\       
         \texttt{td} &  Unfiltered &  30.39 & 35.50 &  17.12 &  49.90 & 0 &  \\
            &  Filtered   &  30.15 & 35.45 &  17.30 &  49.90 & 0 &  6\\ 
              &    &   &  &   &   & \\ 
         \texttt{ct} &  Unfiltered &  33.48 & 37.60 &  15.37 &  50.30 & 0 &  \\
            &  Filtered   &  33.46 & 37.60 &  15.37 &  50.30 & 0 & 1 \\ 
              &    &   &  &   &   & \\         
         \texttt{bb} &  Unfiltered &  30.48 & 34.00 &  15.81 &  50.50 & 0 &  \\
            &  Filtered   &  30.25 & 33.80 &  15.88 &  50.50 & 0 & 3  \\ 
              &    &   &  &   &   & \\             
         \texttt{ml} &  Unfiltered &  28.20 & 31.00 &  15.00 &  50.00 & 0  & \\
            &  Filtered   &  28.19 & 31.00 &  15.00 &  50.00 & 0 & 0  \\ 
            \hline
    \end{tabular}
        \label{tab:filtered_site_stats}
\end{table}
        
    A total of 1,545 (0.30\,\%)  out of 520,582 images were identified as anomalous by the fqi filter.  These made up 9.50\,\% of the total anomalous images found for the testing period. Figure~\ref{monthly_rejects} shows that there were two major periods in August and November in 2014 when most rejections happened. In addition, the filter brought down the duty cycle to zero only for 5 days.  While one of these days was during an expected operational period, many of these days were during a low-performance period during the month of August 2014. This is clearly visible in Figure~\ref{dc_drop_ud} where we display the daily duty cycle for the entire test period. Also, the fqi filter reduced the mean duty cycle of the \texttt{ud} from 19.80\,\% to 19.74\,\% but did not affect the median DC. A comparison between the statistics of filtered and unfiltered images is given in Table~\ref{tab:filtered_site_stats}.      
                
\subsubsection{El Teide (\texttt{td})}
\begin{figure}[t]
            \centering
                \includegraphics[width = \textwidth,height=14.85 cm]{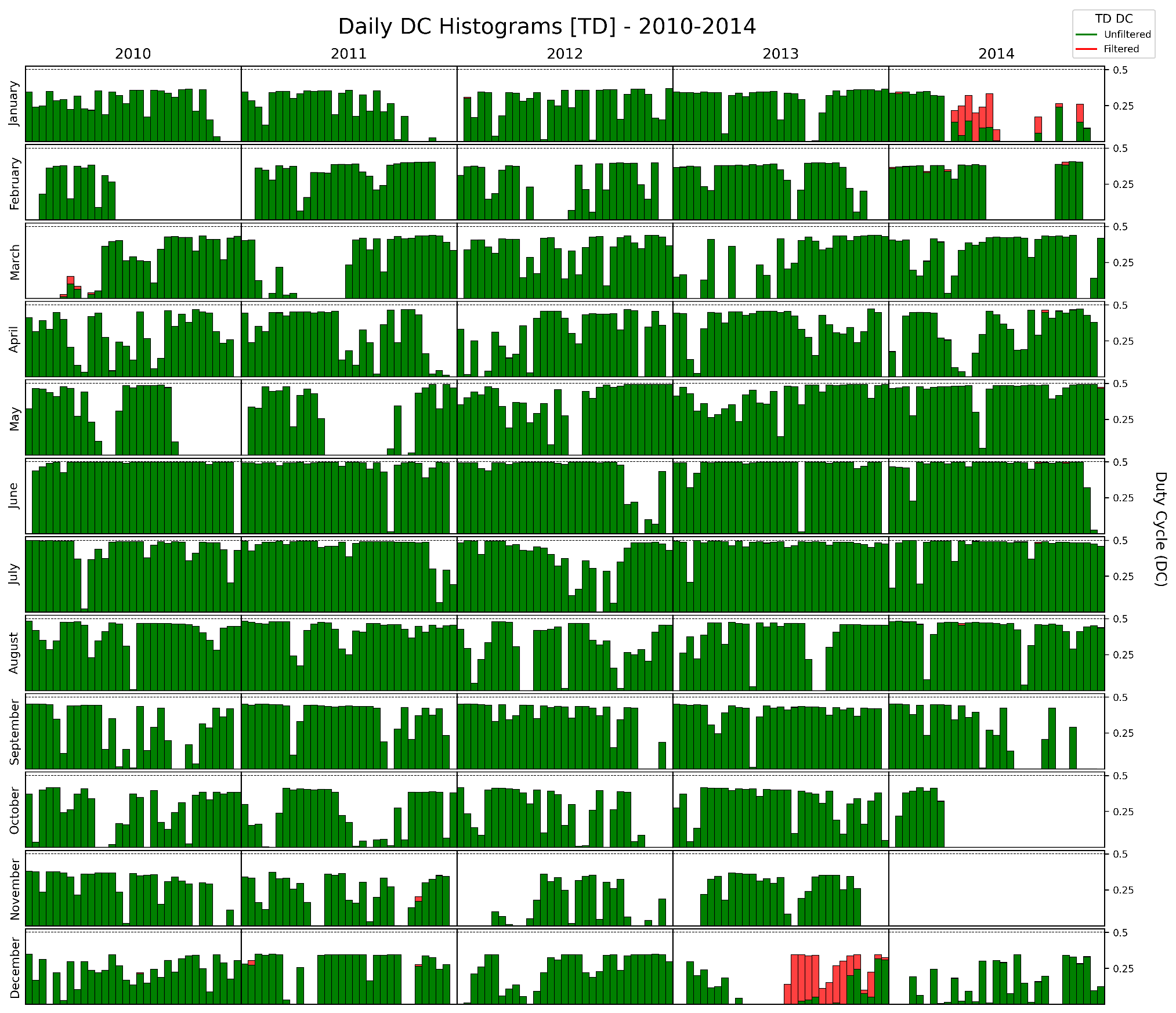}
                     \caption{\it Daily duty cycle for the \texttt{td} site for the testing period from 2010 through 2014. The effects of the fqi filter are shown in red. The significant drop in duty cycles are apparent in December 2013 and January 2014.}
                \label{dc_drop_td}
         \end{figure}

    The El Teide (\texttt{td}) site had a mean and median daily duty cycles of 30.39\,\% and  17.12\,\%, respectively,  for the testing period.
    As exhibited in Figure~\ref{site_dc_hist}, the non-zero duty cycle days for the \texttt{td} site were disproportionately distributed towards 0.50 with most of the observing days being reported as above a duty cycle of 0.30 and nearly 25\,\% of reported days being between 0.45 and 0.50. While \texttt{td} does not experience the seasonal drops in observations similar to \texttt{ud}, it provided good coverage during the months of April  to September (Figure~\ref{dc_drop_td}). As summarized in Table~\ref{tab:simultaneous_obs}, the statistical analysis of the total observations for 799,160 minutes at \texttt{td} suggests that it collected independent observations for 24.05\,\% of those minutes, and the simultaneous observations with one, two and three other sites for 46.81\,\%, 23.49\,\% and 5.64\,\%, respectively.  
    From Figure~\ref{site_combos}, we notice that the \texttt{td} site was most often observed with the \texttt{ct} and \texttt{bb} sites.\\
 
    From Table~\ref{tab:site_contrib}, we note that the implementation of fqi filter rejected  6510 observations. These rejected images accounted for 0.81\,\% of the total observations at \texttt{td} and  42.51\,\% of the all anomalous images identified by the fqi network, making the \texttt{td} site the largest contributor of anomalous images for the testing period.  As a result of the filter's application, the duty cycles for 6 days   completely reduced to 0 (Table~\ref{tab:filtered_site_stats}). It is also observed from Figure~\ref{monthly_rejects} that the vast majority of the anomalous images were generated during   December 2013 through February 2014. The application of the fqi filter reduced the mean duty cycle of the \texttt{td} site from 30.39\,\% to 30.15\,\%, however increased the median duty cycle from 17.12\,\% to 17.30\,\%. 
      
    \begin{figure}[t]
        \centering
            \includegraphics[width = \textwidth,height=14.85 cm]{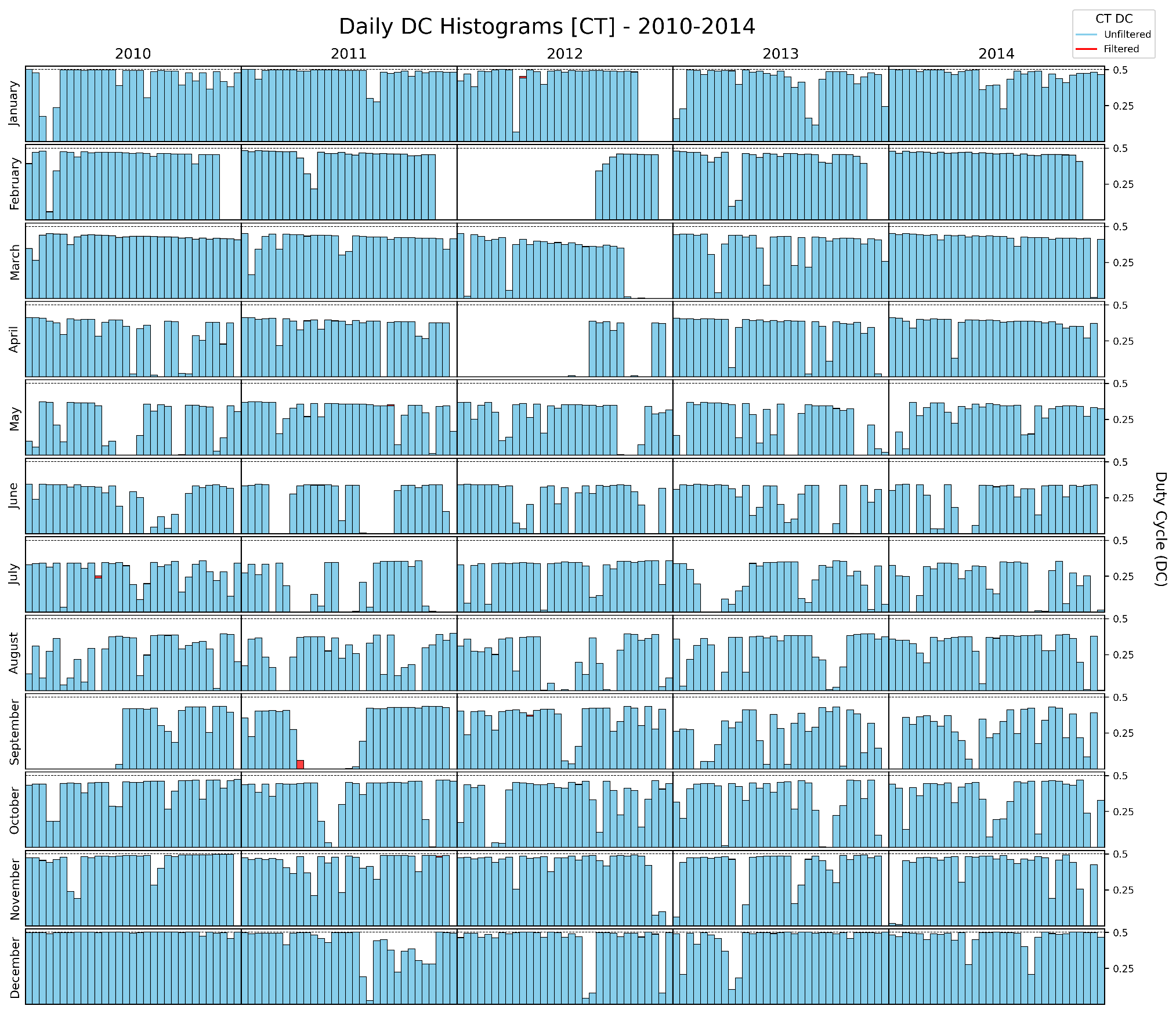}
                \caption{\it Daily duty cycle for the \texttt{ct} site for the testing period from 2010 through 2014. The effects of the fqi filter are shown in red. The drop in the duty cycle is apparent in on one day September 2011 where the duty cycle dropped to 0.}
            \label{dc_drop_ct}
    \end{figure}

 \subsubsection{Cerro Tololo (\texttt{ct})}
    The Cerro Tololo (\texttt{ct}) site had an average daily duty cycle of 33.76 \,\% and a median duty cycle of 37.60\,\% for the testing period. 
     It is demonstrated in Figure~\ref{site_dc_hist} that the non-zero duty cycle days for the \texttt{ct} site were largely above 0.30 with the majority of days reporting a duty cycle between 0.45 and 0.50. During the testing period, the \texttt{ct} site produced consistent observations but with some notable lapses in September of 2010 and 2011, and February and April of 2012. These gaps are clearly visible in Figure~\ref{dc_drop_ct}. The application of the fqi filter resulted in one day only having its duty cycle reduced entirely to zero.  Of the 880,258 images collected at \texttt{ct}, 328 (0.04\,\%) of those images were identified as anomalous by the neural network (Table~\ref{tab:site_contrib}), making it the lowest contributing site to the anomalous images and accounted for 2.14\,\% of the total anomalous images found during the testing period. \\ 

     Of the 880,258 minutes observed by \texttt{ct} during the testing period, it observed independently for 15.57\,\% of the total minutes. In addition, its observations were overlapped with one, two and three other sites for 44.50\,\%,  34.71\,\%, and  5.21\.\% of the total minutes, respectively. It is also seen from Figure~\ref{site_combos} that the \texttt{ct} site observed  most frequently with the \texttt{td} and \texttt{bb} sites. The application of the fqi filter reduced the mean duty cycle for \texttt{ct}  from 33.48\,\% to 33.46\,\% and did not affect the median duty cycle.  
    \begin{figure}[t]
        \centering
             \includegraphics[width = \textwidth,height=14.85 cm]{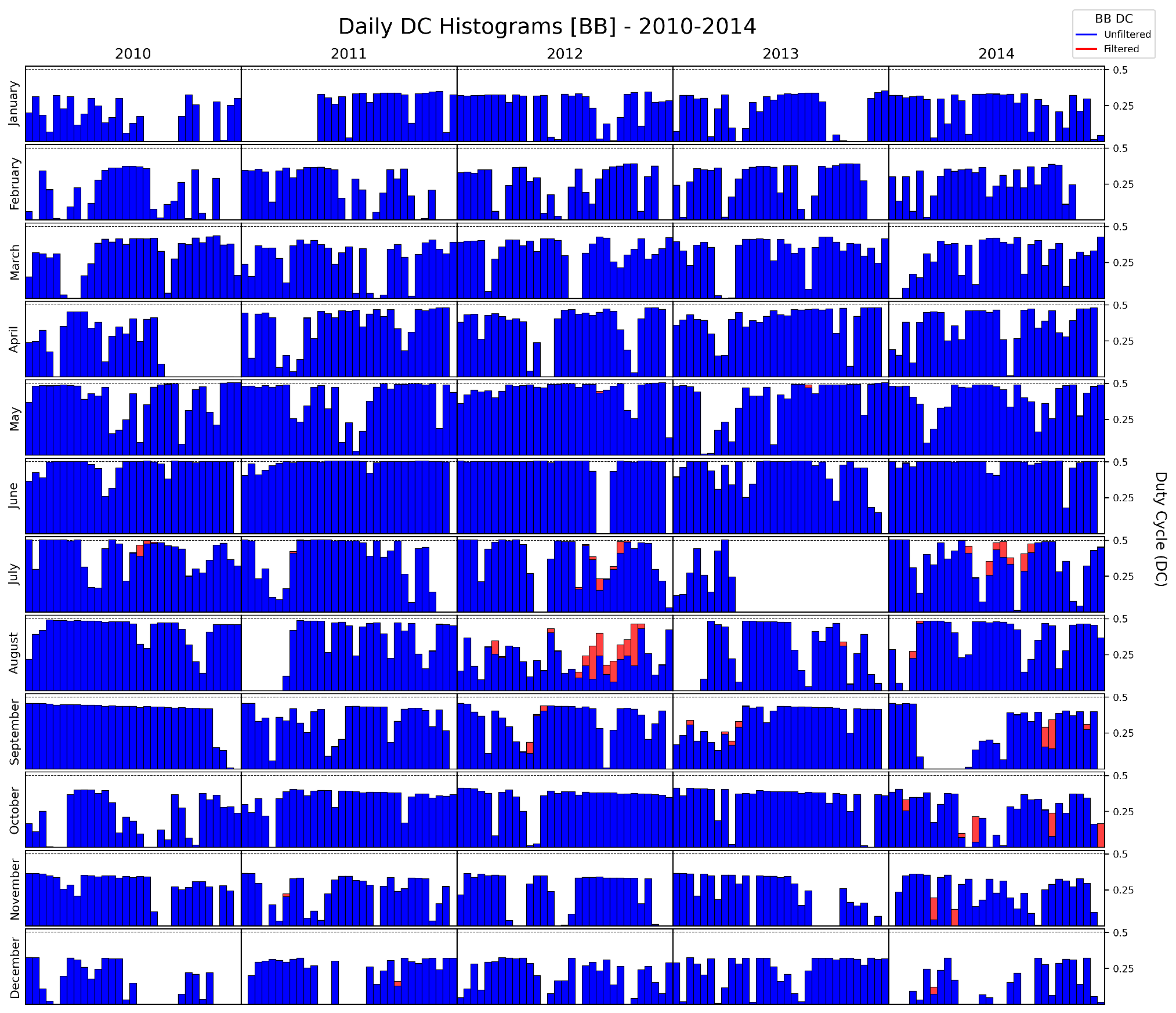}
                \caption{\it Daily duty cycle for the \texttt{bb} site for the testing period from 2010 through 2014. The effects of the fqi filter are shown in red. The drops in duty cycles are apparent in July 2010, July -- September 2011, August -- September 2013 and July -- December 2014.}
            \label{dc_drop_bb} 
    \end{figure}

\subsubsection{Big Bear (\texttt{bb})}
     The average daily duty cycle of Big Bear (\texttt{bb}) site was 30.48\,\% with a median duty cycle of 34.00\,\%. 
     As shown in Figure~\ref{site_dc_hist}, a little over 10\,\% of the reported days for the testing period were reported as having a duty cycle of 0, while 20\,\% of the reported days had a duty cycle between 0.45 and 0.50 for the \texttt{bb} site. From Figure~\ref{dc_drop_bb}, we find that the \texttt{bb} site provided relatively good coverage  in the months of May and June. Of the 801,498 images from \texttt{bb}, 5974 (0.75\,\%) of them were rejected for being anomalous (Table~\ref{tab:site_contrib}). This comprised 39.01\,\% of all anomalous images found for the testing period, making the \texttt{bb} site the second largest contributor of the erroneous images in the network. There were two major events, as shown in Figure~\ref{monthly_rejects},  which accounted for the majority of the rejected images from the \texttt{bb} site, the first occurring between July and September of 2012, and the second occurring between July and November of 2014. It is important to mention that the RMS filter became operational in July 2014 as discussed in Section~\ref{Current_QAQC}, but it did not exclude all the anomalous images after its implementation and allowed these images to get ingested in downstream pipelines and related data products. The application of the fqi filter on the \texttt{bb} site images resulted in reducing the duty cycle to 0 for 3 days as listed in Table~\ref{tab:filtered_site_stats}. \\

                \begin{figure}[t]
                    \centering
                        \includegraphics[width = \textwidth,height=14.85 cm]{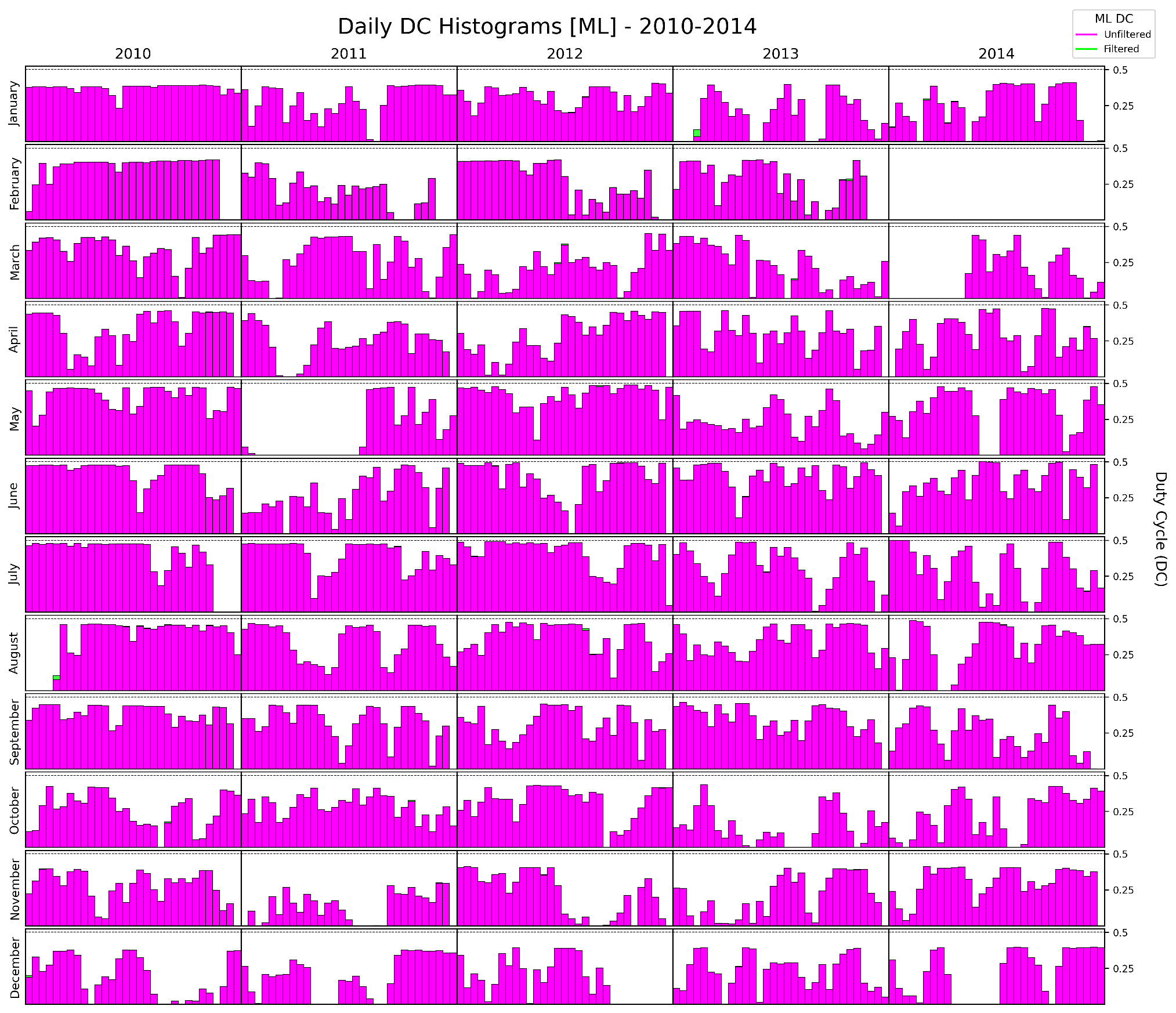}
                            \caption{\it Daily duty cycle for the \texttt{ml} site for the testing period from 2010 through 2014. The effects of the filter are shown in green. There were no major drops in the duty cycle due to the applications of fqi filter, however one can notice changes in August 2010 and January 2013. } 
                            \label{dc_drop_ml}
                \end{figure}
       
    Of the 801,498 minutes observed at the \texttt{bb} site during the testing period, it provided independent observations for 10.14\,\% of those minutes, and was one of two, three and four sites observed at the same time  for 40.78\,\%, 43.31\,\% and  5.77\,\% of the total minutes, respectively  (Table~\ref{tab:simultaneous_obs}). It is noticed from Figure~\ref{site_combos} that the \texttt{bb} site most often observed with the \texttt{ml} and \texttt{ct} sites. Due to the implementation of the fqi filter, the mean daily and median duty cycles were reduced from 30.48\,\% to 30.25\,\%, and  34.00\,\% to 33.80\,\%, respectively. 

\subsubsection{Mauna Loa (\texttt{ml})}   
    The last site to be considered is Mauna Loa (\texttt{ml}). The daily duty cycles for the testing period are shown in Figure~\ref{dc_drop_ml}. It had an average daily duty cycle of 28.20\,\% with a median duty cycle of 31.00\,\%.  As presented in Figure~\ref{site_dc_hist}, the \texttt{ml} site had a DC of 0 for a little over 10\,\% of the reported days with the majority of its days being reported above 0.30. For the testing period, \texttt{ml} had relatively consistent performance except for two large outages occurring in May of 2011, and February/March of 2014. Additionally, observations collected during  October through December were relatively less due to its geographic location. Nevertheless, this site has crucial importance because of its location in the middle of the Pacific ocean and provides vital support for overlapping observations to its neighboring sites. Of the 741,636 images collected by the \texttt{ml} site, 425 (0.06\,\%) of these images were identified as anomalous by the network (Table~\ref{tab:site_contrib}). These comprised 2.78\,\% of all the anomalous images found during the testing period. These images were rejected during multiple small epochs as seen in Figure~\ref{monthly_rejects} and the implementation of the fqi filter did not result in the duty cycle of any day being brought down to zero for this site (Table~\ref{tab:filtered_site_stats}). \\
       
   The \texttt{ml} site collected observations for 741,636 minutes in total and  11.07\,\% of them were independent observations. As given in Table~\ref{tab:simultaneous_obs},  45.9\,\% of its observations were overlapped by one other site, 36.8\,\%  and 6.2\,\% were overlapped with two and three other sites, respectively. An examinations of Figure~\ref{site_combos} reveals that the \texttt{ml} site had maximum overlap with \texttt{bb} and then with \texttt{ct} sites.  Due to the implementation of the fqi filter on \texttt{ml} site images, the mean daily duty cycle  reduced from 28.20\,\% to 28.19\,\%, and the median duty cycle remained unchanged at 31.00\,\%. 
                
\section[{Case Studies for Farside Map Effects}]{Case Studies for Farside Map Effects}

\subsection{Noise Reduction in Merged \texttt{fqi} Images}
        In order to illustrate the impact of single image exclusions on merged \texttt{fqi} images, we have selected three examples from the database where fqi filter identified erroneous site images. These were taken at 20100805t2005, 20141031t1742, and 20140625t0502, representing examples of some of the common types of anomalies seen in the site \texttt{fqi} images. 

    \begin{enumerate}
        
    \item{\textbf{20100805t2005}}
            The first example includes a fully contaminated site image from \texttt{ml} shown in the top left panel of Figure~\ref{cleanmrfqi_2}. These types of artifacts often start at one corner of the site images and progress until they cover the entire observed hemisphere. There were two more sites, \texttt{ct} and \texttt{bb}, that provided observations at the same time.  When anomalous \texttt{ml} image was combined with good \texttt{ct} and \texttt{bb} images, it contaminated the merged product (\texttt{mrfqi}) shown in the top right panel of the same figure, completely erasing the valuable Doppler signals. The application of fqi filter correctly identified \texttt{ml} image as anomalous and rejected its inclusion in the merged image.  By eliminating the anomalous image, shown in the bottom row, we obtained a good quality merged \texttt{mrf6i} image that can be reliably used in the farside pipeline or any other scientific study.
                \begin{figure}[H] 
                \centering    
                    \includegraphics[width = 0.79\textwidth]{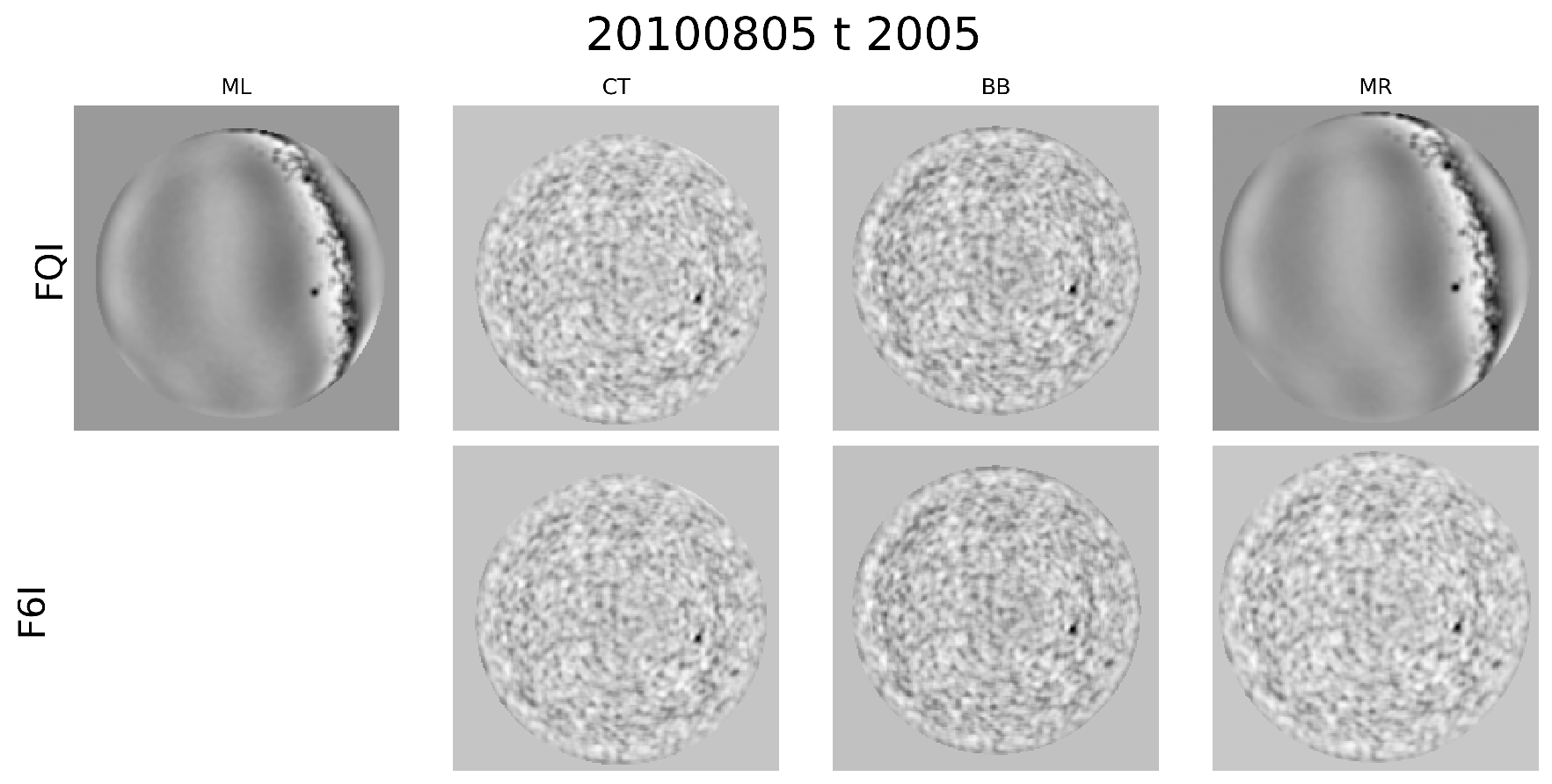}
                        \caption{\it Effects of including (top row) and excluding (bottom row) anomalous site data in merged \texttt{fqi} image for 20100805t2005. The  \texttt{ml} image was identified by the fqi filter network. The merged images without contaminated data are labeled F6I instead of FQI for the traditional pipeline.}
                    \label{cleanmrfqi_2}
                   \vspace{-.25cm}
                \end{figure}

    \item{\textbf{20141031t1742}}
        Another example shows a corrupted image with the so-called dinner plate artifacts. These artifacts are a ring pattern which often appear from time to time and disappear suddenly. This is a more common type of artifact seen in the site images that may affect either full or half disk. Its origin is not yet fully understood, however it adversely affects the quality of the merged images.  As visible in Figure~\ref{cleanmrfqi_1}, the rings in \texttt{bb} image translated to the final merged image, introducing subdued noise in the ring pattern. The fqi filter eliminated the \texttt{bb} image from the list of images to be merged for this timestamp, thus allowed us to produce a merged image without any artifact.

                \begin{figure}[H] 
                    \centering    
                        \vspace{-0.5\intextsep}
                         \includegraphics[width = 0.79\textwidth]{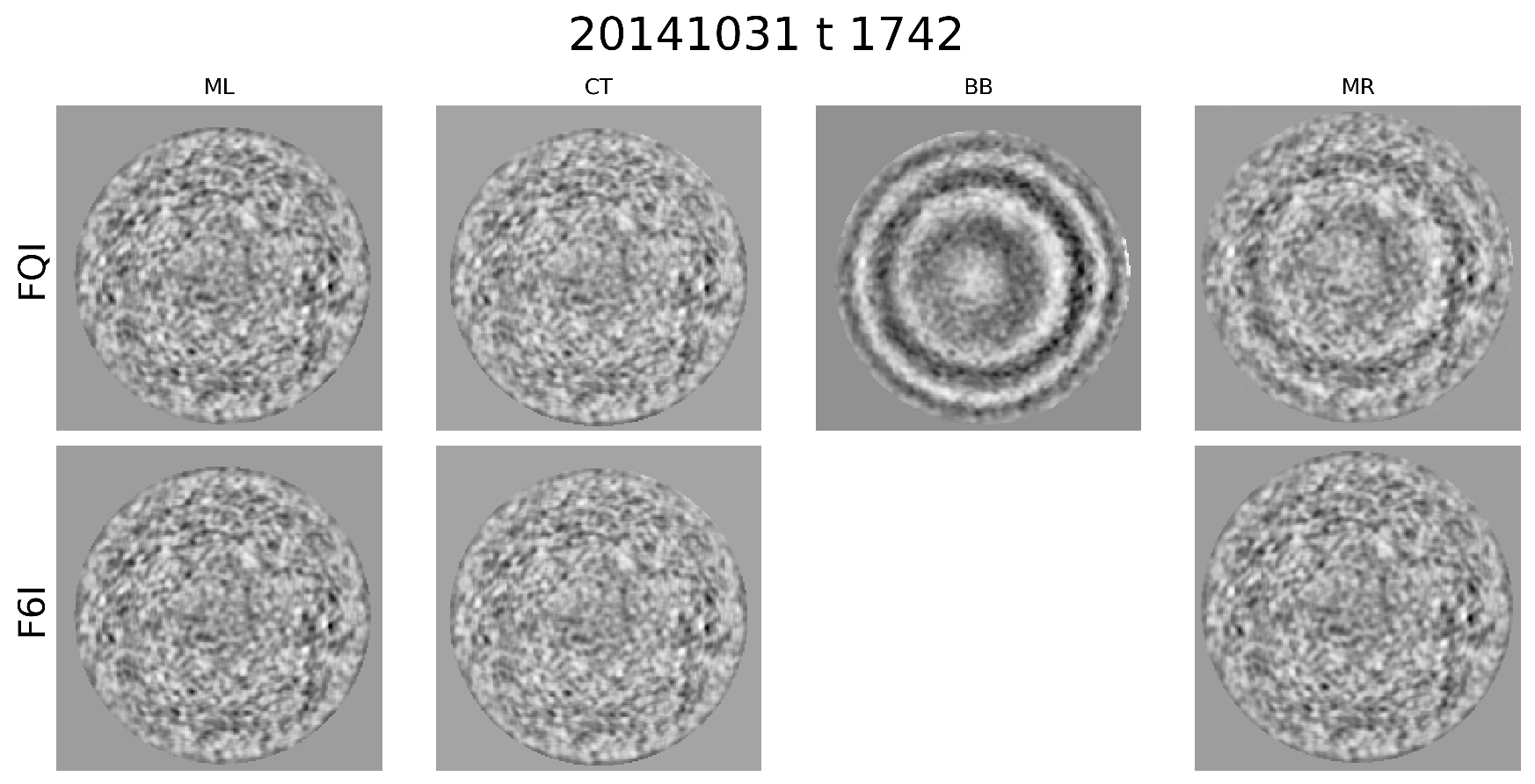}
                         \caption{ \it Same as Figure~\ref{cleanmrfqi_2}, but for the merged image at 20141031t1742 excluding the \texttt{bb} image.}
                        \label{cleanmrfqi_1}
                    \end{figure}

     \item{\textbf{20140625t0502}}
        The third example of the contaminated site \texttt{fqi} image includes a pressed plate artifact. These often appear suddenly and can completely erase the data from good images when included in merged (\texttt{mrfqi}) images. Due to their significant impact, it is important for the network to identify and exclude these images, as they can introduce significant noise from a single point. By excluding such artifacts, we are able to produce clean \texttt{mrfqi} images. Figure~\ref{cleanmrfqi_3} shows an image from \texttt{le} with this type of artifact. Although \texttt{ud} image at the same time was absolutely normal, the inclusion of abnormal \texttt{le} produced corrupted merged image. The result of using the fqi filter for this timestamp as shown in the bottom row.
        \begin{figure}[H] 
        \centering    
            \vspace{-0.5\intextsep}
        
            \includegraphics[width = 0.79\textwidth]{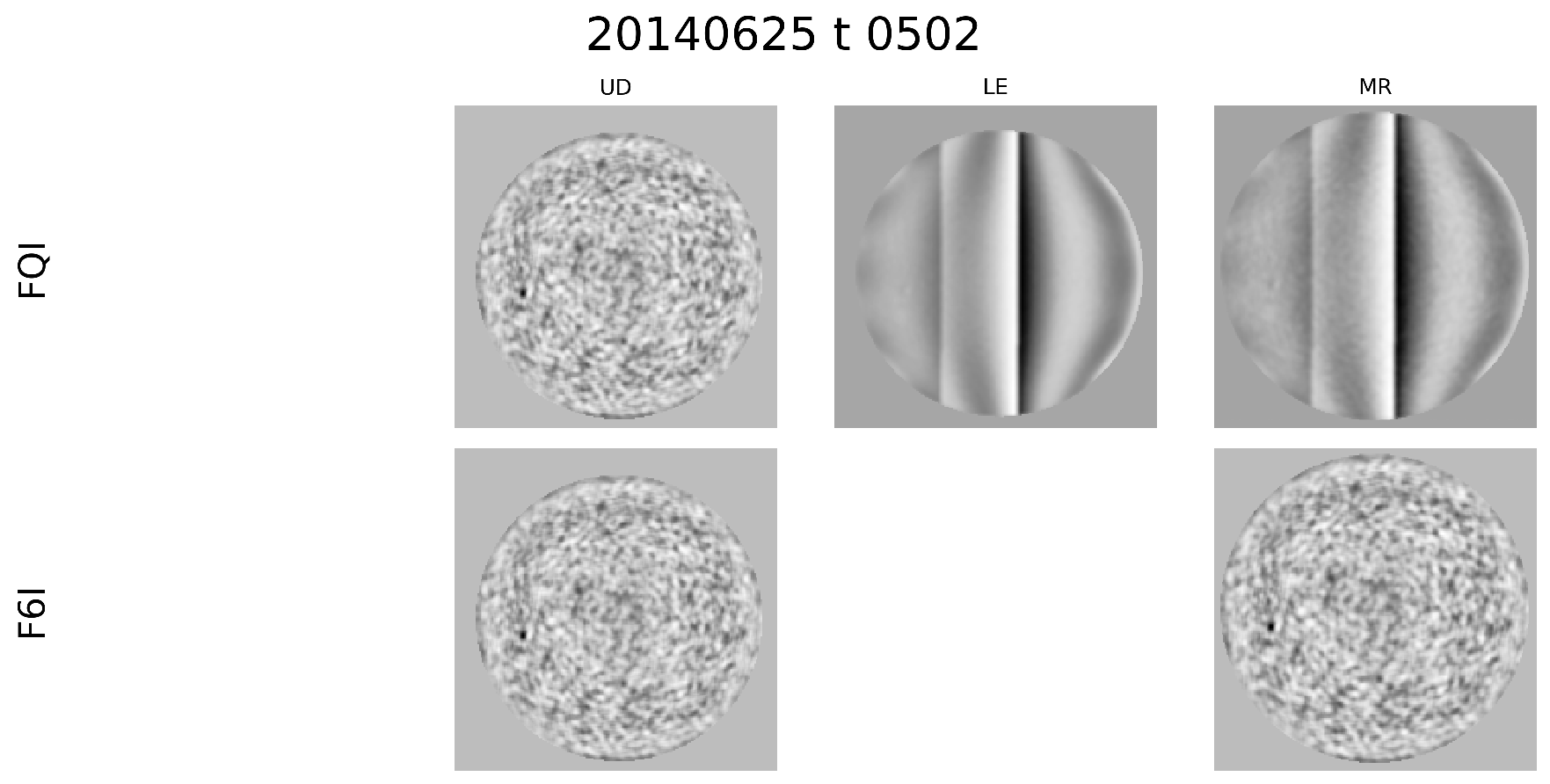}
                \caption{\it Same as Figure~\ref{cleanmrfqi_2}, but for the merged image at 20140625t0502 excluding the \texttt{le} image.}  
                \label{cleanmrfqi_3}
                \vspace{-0.5\intextsep}
        \end{figure}
\end{enumerate}
    \subsection[{Noise Reduction in Farside Maps}]{Noise Reduction in Farside Maps}
        While the effects of the fqi network are not visible in every farside map, it is more prominent during the periods when the fqi filter identified several anomalous images and the seismic maps were extremely noisy. The network filter significantly reduced the noise in exchange of a small drop in duty cycle. In order to demonstrate the efficacy  of the small number of images excluded by the fqi filter, we discuss three examples of the farside seismic maps at 1200 UT on 20111126, 20111202, and 20120206.  These maps are calculated using 1440 merged images. The statistical effects of the implementation of the filter on farside maps are summarised in Table~\ref{tab:map_samples}.
      
        \begin{table}[H]      
            \vskip 0.15 cm
           \caption{\it Summary statistics for production (\texttt{fqm}) and cleaned (\texttt{f6m}) sample phase shift maps.  }
             \centering
              \label{tab:map_samples}
            \begin{tabular}{ccccccccc}\hline
                     Date&  Map &  Number of & Duty & \multicolumn{3}{c}{Phase Shift (radians)} &  Standard  & Signal-to-noise \\ \cline{5-7}
                         &  Type& Input Images & Cycle & Min&  Max&  Mean                &  Deviation & Ratio \\ \hline
                     20111126t1200&  \texttt{fqm}&  1395 & 0.97 & $-$3.34&  2.738&  0.025&  0.401& 0.063 \\
                                  &  \texttt{f6m}&   1383 & 0.96 & $-$0.206&  0.133&  $-$0.017&  0.051& 0.338 \\
                                  & & & & & &\\
                     20111202t1200&  \texttt{fqm}& 1430 &  0.99   & $-$0.691&  0.973&  0.027&  0.173& 0.155 \\
                                  &  \texttt{f6m}&   1430 & 0.99  & $-$0.2&  0.134&  $-$0.015&  0.050& 0.307\\
                     & & & & & &\\
                     20120206t1200&  \texttt{fqm}& 1239 & 0.86 & $-$0.594&  1.064&  0.049&  0.155& 0.318\\
                     &  \texttt{f6m} &1235 &  0.86 & $-$0.241&  0.185&  $-$0.003&  0.055& 0.054
                     \\
                     \hline
            \end{tabular}    
            \vskip 0.25 cm
        \end{table}
\begin{enumerate}

\item {\textbf{20111126t1200}}
        The first example is the farside seismic map with the timestamp of 20111126t1200. This map was initially computed with  44 contaminated site \texttt{fqi} images from the \texttt{td} site and the exclusion of these images resulted in the loss of 12 merged \texttt{fqi} images. The unfiltered images introduced significant noise in the farside as displayed in the left panel  of Figure~\ref{cleaned_1}, however the barring these images not only resulted in a striking visual improvement in clarity (right panel in the same figure), but also a statistically noticeable improvement.  From Table~\ref{tab:map_samples}, it can be seen that  the standard deviation dropped from 0.401 to 0.051 radians, indicating a far more narrow band of readings. Additionally, the minimum and maximum values changed from $-$3.34 to  $-$0.206 radians and 2.738 to 0.133 radians, respectively. The values with unfiltered images were excessive and had no physical significance. The mean being reduced from 0.025 to $-$0.017 radians is more in line with the predominantly negative phase shift values that we expect to see from the helioseismic maps. The signal-to-noise ratio also increased significantly from 0.063 to 0.338. 

             \begin{figure}[h!] 
                \centering    
                  \includegraphics[width = 0.93\textwidth]{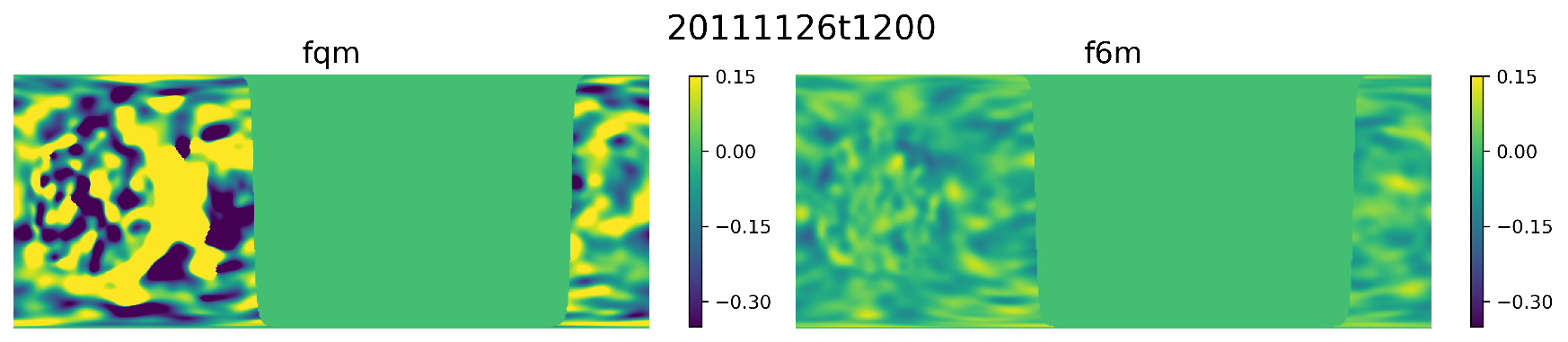}
                    \caption{\it Farside phase-shift maps using contaminated (\texttt{fqm}) and cleaned (\texttt{f6m}) merged site images for 20111126t1200. This \texttt{fqm} map was contaminated by anomalous site images which were identified by the fqi network. The scale depicts the phase-shift values in radians with the frontside of the sun masked out in green.}
                \label{cleaned_1}
            \end{figure}
         
\item {\textbf{20111202t1200}}
        The second example is the farside seismic map at 20111202t1200. This map, displayed in Figure~\ref{cleaned_2},  also has visible and statistical  improvements when filtered images are used. The contaminated 83 images from the \texttt{le} and \texttt{ml} sites produced a noisy map as shown in the left panel and the exclusion of which did not reduce the number to good  merged \texttt{fqi} images. Similar to previous case, the exclusion of erroneous images dropped the standard deviation from 0.173 to 0.050 radians, indicating a more accuracy in the farside map.   The minimum and maximum values also modified from $-$0.691 to $-$0.200  radians and from 0.973 to 0.134 radians, respectively. The modified values are within the expected range of phase shifts 
        and the mean changed from 0.027 to $-$0.015 radians. Again, this shows the predominantly negative phase shift values that we expect from the helioseismic maps. The signal-to-noise ratio again doubled  from 0.155 to 0.307 (Table~\ref{tab:map_samples}). 
     \begin{figure}[H] 
        \centering
        \includegraphics[width = 0.93\textwidth]{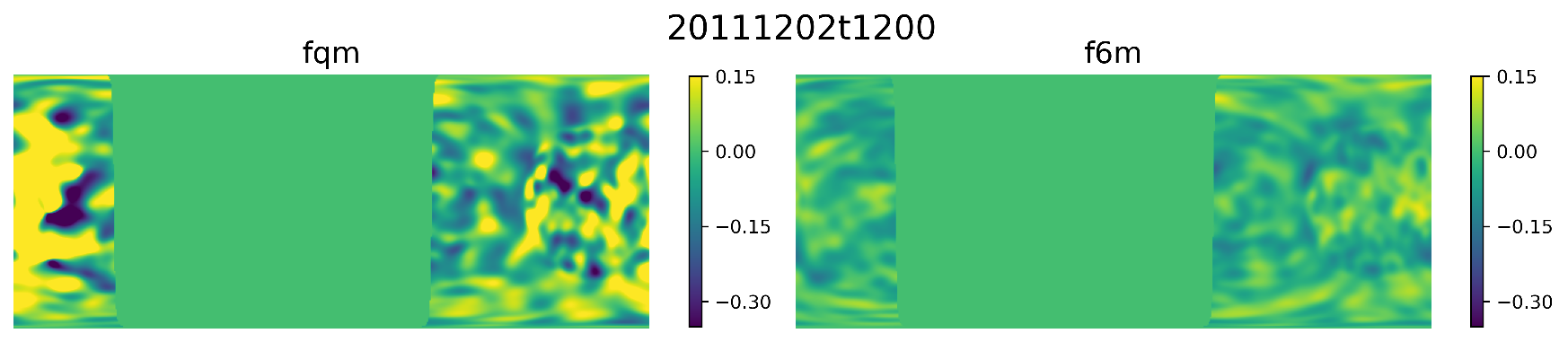}
            \caption{\it Same as Figure~\ref{cleaned_1}, but for 20111102t1200.}
        \label{cleaned_2}
    \end{figure}
 
\item {\textbf{20120206t1200}}
        The last  illustration is the map from 20120206t1200 shown in Figure~\ref{cleaned_3}. This is the most interesting map because it appears to be a seismic maps for the abnormally high-activity period at first glance. However, a close inspection of the images used to produce this map reveals that the \texttt{fqm} map was contaminated by 44 anomalous images from the \texttt{td} site and the exclusion of these images resulted in the loss of four merged \texttt{fqi} images. The result of the exclusion of these images produced a map with far more consistency that lost large phase-shift values which it initially displayed. The maximum phase-shift value was reduced from 1.064 to 0.185 radians, while it's minimum was also brought in from $-$0.594 to $-$0.241 radians. This also reduced the standard deviation from 0.155 to 0.055 radians. However, the signal-to-noise ratio reduced from 0.318 to 0.54 (see Table~\ref{tab:map_samples}). This map is an excellent example of the necessity of the fqi neural network filter. While it clearly had anomalous images as its components, statistical analysis of the map itself would not have revealed it to be faulty. 
        \begin{figure}[H] 
            \centering    
             \includegraphics[width = 0.93\textwidth]{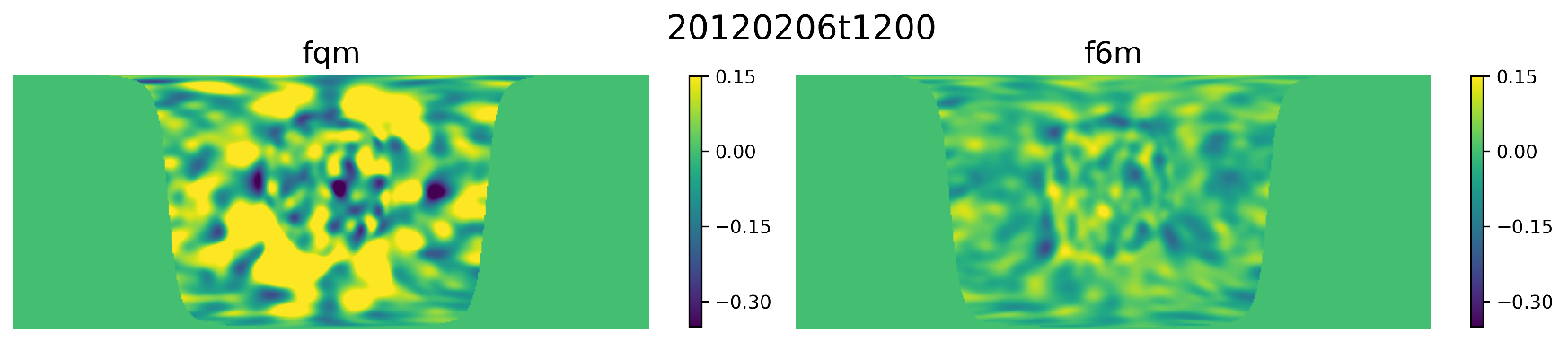}
                \caption{\it Same as Figure~\ref{cleaned_1}, but for 20120206t1200.}
             \label{cleaned_3}
        \end{figure}
\end{enumerate}
                
    These examples show the adaptability of the discriminatory network and its ability to provide quality assurance and control measures in a variety of statistically novel situations. We further checked the credibility of \texttt{f6m} maps by comparing them with those produced by both Helioseismic and Magnetic Imager  farside pipelines based on the helioseismic holography (a similar technique used by the NISP farside pipeline) and the time-distance techniques, and found a close correspondence between them. It should be noted that HMI has several advantages being in space and the images are less prone to any artifacts.

\section[{Network Updates and Improvements}]{{Network Updates and Improvements}}
\label{Updates}
    There are two methods through which the performance of the current fqi network can be improved. The first of these is re-training, during which the network weights would be updated through training on newly labeled data. The second of these methods is a network re-design, in which the current network architecture would be replaced by an updated architecture design.  

\subsection[{Re-training}]{Re-Training}
        In order to make network re-training and updates as easy as possible, a Jupyter notebook has been created and uploaded to the NISP Data Center's GitLab under the name \texttt{fqi\_Network\_Retraining}. While the README file in the repository contains specific step-by-step instructions, the seven steps of implementing new weights are also outlined in this document. This explanation will assume that the user has already locally cloned the retraining repository and has the appropriate environmental packages installed. 
    
\begin{enumerate}
     \item \textbf{Load the newly labeled data:} Load the newly labeled normal and anomalous data into \texttt{.lst} files so that it can be included in the training regime. The notebook assumes that a number of images which have not been previously used in training the network have been manually labeled into Normal and Anomalous classifications. Directories to the new anomalous and normal data should be collected and stored in two files named \texttt{anom\_new.lst} and \texttt{norm\_new.lst} respectively. These should be stored in the \texttt{data/new\_data/} subdirectory of the retraining regime. The re-training regime is designed to take directories from the \texttt{oQR} directory. These files will be combined with old training data in order to update the fqi network weights. At the end of the re-training process, there is an opportunity to combine the new and old datasets so that the process can be repeated as many times as possible. The more data that is added to the re-training data, the better the updated results will be. It is recommended that re-training be conducted with at least a 1\,\% expansion on the current dataset, but more is always better. 
        
    \item \textbf{Ensure old network weights are available:} Ensure that the old network weights are available. These are stored as the \texttt{.pt} file from which the fqi Network loads its weights in deployment. This file needs to be stored in the \texttt{params/old\_params} directory, and does not need to have a specific name as the re-training regime will pick up any file stored in this directory. 
        
    \item \textbf{Run the training regime:} Run the training regime. So long as the environment is properly set up, this should only entail running the Jupyter notebook. All of the required packages are stored in the first cell of the notebook. 
        
    \item \textbf{Evaluate the network performance:} Evaluate the network performance. New network performance can be seen at the bottom of the Jupyter notebook, where the network statistics, confusion matrix, and a visual sample of said matrix results can be seen. If there are patterns of images which the network has trouble identifying, then the suggestion is to add more of those images to the new dataset and re-run the program before re-deploying. It should be noted that the goal is not absolute perfection, as this would be indicative of over-training and may, counter-intuitively, lead to poor performance in deployment, but rather the minimization of false-negative results so as to avoid the taking of large sections of high-activity periods. 
        
    \item \textbf{Make adjustments to the network:} If network performance is not sufficiently improved, make adjustments. These include gathering more data, adjusting hyperparameters such as learning rate, batch size, or number of epochs, and double-checking the sample for incorrectly labeled images. 
        
    \item \textbf{Update the network weights:} Update the network weights. The new weights will be stored in \texttt{params/new\_params/Spotty\_Net\_Update.pt}. This file can be renamed in accordance with the decided naming conventions for the fqi Network weights and replaced in the appropriate file. 
        
    \item \textbf{Merge the new and old data:} Once the network performance is satisfactory, merge the new and old data to prepare the re-training regime for future updates. Replace the files \texttt{anom\_new.lst}, \texttt{anom\_old.lst}, \texttt{norm\_new.lst}, \texttt{norm\_old.lst} with the files found under \texttt{new\_data} of the same name. After this is done, you may delete the \texttt{new\_data} directory. 
\end{enumerate}
    
\subsection[{Network Design Updates}]{Network Design Updates}
    The second option for improving the performance of the fqi network would be a fundamental re-design. While the specifics of a network re-design are outside the scope of this discussion, it would broadly entail re-training on the existent dataset with an updated network architecture. Such a campaign would be more intensive than a simple weight update and thus should be considered carefully. However, given the rapid pace of improvement in the field of machine learning, new network designs may provide significant performance improvements. New strategies such as GANs or other fundamental improvements in architectures should be considered for performance improvements \citep{dimattia2021survey,xia2022gan-based}. \\
        
    In the event the the network is re-designed, both the weights file and the network architecture files should be replaced. Once these are replaced and the associated calls in the code have been updated, that should be sufficient to run the network on the new architectures.  

\section[{Summary}]{Summary}
    The binary classification network developed and analyzed in this report has proven to be a highly effective method for reducing the impact of anomalous site-images on the noise of NISP's farside helioseismic maps. This neural network-based classification methodology is resilient enough to accurately identify anomalous images despite their statistical variability. It is also more selective than the current thresholding technique, preventing the exclusion of large amounts of valuable high-activity data. The spacial variability of the GONG network's observational sites also provide adequate resiliency to compensate for the majority of anomaly production events at any single site. When we also consider the potential for re-training and redesign, the fqi network has proven to be invaluable and adaptable QAQC methodology for the NISP farside program. We hope that the performance and adaptability 
    of this new methodology serves as a proof of concept for the implementation of similar measures for other NISP data products in the future.\\     
    
\addcontentsline{toc}{section}{\bf Acknowledgements}
    {\bf Acknowledgements:} This work utilises GONG data obtained by the NSO Integrated Synoptic Program, managed by the National Solar Observatory, which is operated by the Association of Universities for Research in Astronomy (AURA), Inc. under a cooperative agreement with the National Science Foundation (NSF) and with contribution from the National Oceanic and Atmospheric Administration (NOAA). The GONG network of instruments is hosted by the Big Bear Solar Observatory, High Altitude Observatory, Learmonth Solar Observatory, Udaipur Solar Observatory, Instituto de Astrof\'{\i}sica de Canarias, and Cerro Tololo Interamerican Observatory. This work is partially supported by the Windows On the Universe (WoU) grant funded by NSF.
    
\addcontentsline{toc}{section}{\bf References}
\bibliographystyle{apalike}


\end{document}